\documentclass[reprint,superscriptaddress,amsmath,amssymb,aps,longbibliography,prl]{revtex4-2}

\usepackage{graphicx}
\graphicspath{ {./figures/}}
\usepackage[dvipsnames]{xcolor}
\usepackage[colorlinks=true,pdfpagelabels=false]{hyperref} 
\hypersetup{urlcolor=blue,linkcolor=violet,citecolor=blue}
\usepackage{physics}
\usepackage{mathtools}
\usepackage{times}
\usepackage{orcidlink}
\usepackage[nameinlink,capitalize]{cleveref}

\definecolor{col_green}{RGB}{38,162,105}
\definecolor{col_poly}{RGB}{255,159,72}
\definecolor{col_dim}{RGB}{38,162,105}
\definecolor{col_mon}{RGB}{255,163,72}
\definecolor{col_link}{RGB}{0,0,0}
\definecolor{col_membrane}{RGB}{38,162,105}

\tikzset{
grid/.style={line width=0.25mm, black!25,densely dotted},
dimer/.style={line width=0.55mm, col_link},
}

\newcommand{\plaqA}{\hspace{0.5mm} \begin{tikzpicture}[scale=0.20,baseline=0.1mm]
 \draw[grid] (0,0) -- (1,0);
\draw[dimer] (1,0) -- ++(90:1);
  \draw[grid] (1,0)++(90:1) -- (90:1);
  \draw[dimer] (0,0) -- ++(90:1);
 \filldraw[col_dim] (0,0) circle (6pt);
  \filldraw[col_dim] (1,0) circle (6pt);
    \filldraw[col_dim] (1,0) ++(90:1) circle (6pt);
    \filldraw[col_dim] (90:1) circle (6pt);
 \end{tikzpicture} \hspace{0.5mm}}

 \newcommand{\plaqB}{  \hspace{0.5mm}\begin{tikzpicture}[scale=0.20,baseline=0.1mm]
 \draw[dimer] (0,0) -- (1,0);
\draw[grid] (1,0) -- ++(90:1);
  \draw[dimer] (1,0)++(90:1) -- (90:1);
  \draw[grid] (0,0) -- ++(90:1);
 \filldraw[col_dim] (0,0) circle (6pt);
  \filldraw[col_dim] (1,0) circle (6pt);
    \filldraw[col_dim] (1,0) ++(90:1) circle (6pt);
    \filldraw[col_dim] (90:1) circle (6pt);
 \end{tikzpicture} \hspace{0.5mm}}

 \newcommand{\parallelograma}{ \hspace{0.5mm} \begin{tikzpicture}[scale=0.20,baseline=0.1mm]
 \draw[line width=0.25mm, black] (0,0) -- (1,0) -- (0,1) -- (-1,1) -- (0,0);
 \end{tikzpicture} \hspace{0.5mm}}

  \newcommand{\parallelogramb}{ \hspace{0.5mm} \begin{tikzpicture}[scale=0.20,baseline=0.1mm]
 \draw[line width=0.25mm, black] (0,0) -- (0,1) -- (-1,2) -- (-1,1) -- (0,0);
 \end{tikzpicture} \hspace{0.5mm}}

 \newcommand{\parallelogramaA}{ \hspace{0.5mm} \begin{tikzpicture}[scale=0.20,baseline=0.1mm]
 \draw[grid] (0,0) -- (1,0);
\draw[dimer] (1,0) -- (0,1);
  \draw[grid] (0,1) -- (-1,1);
  \draw[dimer] (-1,1) -- (0,0);
 \filldraw[col_dim] (0,0) circle (6pt);
  \filldraw[col_dim] (1,0) circle (6pt);
    \filldraw[col_dim] (0,1) circle (6pt);
    \filldraw[col_dim] (-1,1) circle (6pt);
 \end{tikzpicture} \hspace{0.5mm}}

  \newcommand{\parallelogramaB}{ \hspace{0.5mm} \begin{tikzpicture}[scale=0.20,baseline=0.1mm]
 \draw[dimer ] (0,0) -- (1,0);
\draw[grid] (1,0) -- (0,1);
  \draw[dimer ] (0,1) -- (-1,1);
  \draw[grid] (-1,1) -- (0,0);
 \filldraw[col_dim] (0,0) circle (6pt);
  \filldraw[col_dim] (1,0) circle (6pt);
    \filldraw[col_dim] (0,1) circle (6pt);
    \filldraw[col_dim] (-1,1) circle (6pt);
 \end{tikzpicture} \hspace{0.5mm}}

\newcommand{\parallelogrambA}{ \hspace{0.5mm} \begin{tikzpicture}[scale=0.20,baseline=0.1mm]
 \draw[dimer] (0,0) -- (0,1);
\draw[grid] (0,1) -- (-1,2);
  \draw[dimer] (-1,2) -- (-1,1);
  \draw[grid] (-1,1) -- (0,0);
 \filldraw[col_dim] (0,0) circle (6pt);
  \filldraw[col_dim] (0,1) circle (6pt);
    \filldraw[col_dim] (-1,2) circle (6pt);
    \filldraw[col_dim] (-1,1) circle (6pt);
 \end{tikzpicture} \hspace{0.5mm}}

\newcommand{\parallelogrambB}{ \hspace{0.5mm} \begin{tikzpicture}[scale=0.20,baseline=0.1mm]
 \draw[grid] (0,0) -- (0,1);
\draw[dimer] (0,1) -- (-1,2);
  \draw[grid] (-1,2) -- (-1,1);
  \draw[dimer] (-1,1) -- (0,0);
 \filldraw[col_dim] (0,0) circle (6pt);
  \filldraw[col_dim] (0,1) circle (6pt);
    \filldraw[col_dim] (-1,2) circle (6pt);
    \filldraw[col_dim] (-1,1) circle (6pt);
 \end{tikzpicture} \hspace{0.5mm}}

\newcommand{\squarewt}{ \hspace{0.5mm} \begin{tikzpicture}[scale=0.30,baseline=.1mm]
 \draw[line width=0.25mm, black] (0,0) -- (1,0)--(1,1)--(0,1)--cycle;
 \draw[line width=.5mm,purple] (0,0)--(1,0);
  \draw[line width=.5mm,purple] (0,1)--++(1,0);
  \node[purple] at (.5,.5) {\small$\wt$};
 \end{tikzpicture} \hspace{0.5mm}}

\newcommand{\squareone}{ \hspace{0.5mm} \begin{tikzpicture}[scale=0.30,baseline=.1mm]
 \draw[line width=0.25mm, black] (0,0) -- (1,0)--(1,1)--(0,1)--cycle;
 \node at (.5,.5) {\small$1$};
 \end{tikzpicture} \hspace{0.5mm}}

\newcommand{\Z}{\mathbb{Z}}

\newcommand{\C}{\mathbb{C}}

\newcommand{\dconfig}{\mathcal C}

\newcommand{\wt}{\alpha}

\begin{document}

\title{Exactly Solvable Topological Phase Transition in a Quantum Dimer Model}

%\tikzexternaldisable
\author{Laura Shou\,\orcidlink{0000-0001-8624-8063}}
\affiliation{Joint Quantum Institute, Department of Physics, University of Maryland, College Park, MD 20742, USA}
\author{Jeet Shah\,\orcidlink{0000-0001-5873-8129}}
\affiliation{Joint Quantum Institute, Department of Physics, University of Maryland, College Park, MD 20742, USA}
\affiliation{Joint Center for Quantum Information and Computer Science, NIST/University of Maryland,
College Park, MD 20742, USA}
\author{Matthew Lerner-Brecher}
\affiliation{Department of Mathematics, Massachusetts Institute of Technology, Cambridge, MA 02139, USA}
\author{Amol Aggarwal\,\orcidlink{0000-0002-8091-2215}}
\affiliation{Department of Mathematics, Stanford University, Stanford, CA 94305, USA}
\author{Alexei Borodin\,\orcidlink{0000-0002-2913-5238}}
\affiliation{Department of Mathematics, Massachusetts Institute of Technology, Cambridge, MA 02139, USA}
\author{Victor Galitski}
\affiliation{Joint Quantum Institute, Department of Physics, University of Maryland, College Park, MD 20742, USA}

\begin{abstract}
We consider a family of generalized Rokhsar-Kivelson (RK) Hamiltonians, which are reverse-engineered to have an arbitrary edge-weighted superposition of dimer coverings as their exact ground state at the RK point. We focus on a quantum dimer model on the triangular lattice, with doubly periodic edge weights. For simplicity we consider a $2\times1$ periodic model in which all weights are set to one except for a tunable horizontal edge weight labeled $\wt$. We analytically show that the model exhibits a continuous quantum phase transition  at $\wt=3$, changing from a topological $\Z_2$ quantum spin liquid ($\wt<3$) to a columnar ordered state ($\wt>3$).  The dimer-dimer correlator decays exponentially on both sides of the transition with the correlation length $\xi\propto1/|\wt-3|$ and  as a power-law at criticality. The vison correlator exhibits an exponential decay in the spin liquid phase, but becomes a constant in the ordered phase, which we explain in terms of loop statistics of the double-dimer model. Using finite-size scaling of the vison correlator, we extract critical exponents consistent with the 2D Ising universality class.   Additionally, we analytically show that the topological R\'enyi entropy of order $\infty$ (topological min-entropy) changes from $\log2$ for the quantum spin liquid phase $\wt<3$, to $0$ for the ordered phase $\wt>3$, thereby analytically confirming the topological nature of the phase transition.
\end{abstract}

\maketitle

\textit{Introduction}--- Quantum spin and quantum dimer models~\cite{rokhsar1988superconductivity,moessner2011introduction,moessner2001resonating,moessner2001phase,misguich2008quantum,ioselevich2002groundstate,schlittler2017phase} provide complementary descriptions of strongly correlated phases of matter. 
In many settings there is an explicit dictionary between the two languages~\cite{balents2002fractionalization,balasubramanian2022classical,balasubramanian2024interplay,shah2025quantumpenrose}, with spin-singlet coverings mapped to hardcore dimers on an auxiliary lattice. 
Within this framework one encounters a rich landscape of phases, ranging from ordered states to a zoo of quantum spin liquids~\cite{savary2017quantum,hermele2004pyrochlore,balents2002fractionalization}. 
The latter have been an active topic of research since Anderson's proposal of a resonating valence-bond state in the context of high-$T_c$ cuprates~\cite{moessner2003ising}, and were later placed in a broader theoretical setting by Kitaev, Wen, and others, linking quantum spin liquids to topics as diverse as frustrated magnetism, emergent gauge theories, and topological quantum error-correcting codes~\cite{kitaev2006anyons,kitaev2003faulttolerant,wen2002quantum,wen2007quantum,levin2005stringnet,moessner2001shortranged,senthil2001fractionalizationPRB,senthil2001fractionalizationPRL,misguich2002quantum,essin2013classifying}.
Moreover, quantum dimer models can be realized experimentally using platforms such as Rydberg atom arrays~\cite{semeghini2021probing,verresen2021,shah2025quantum,zeng2025quantum,yan2022triangular}. 

\begin{figure*}[htb]
\includegraphics{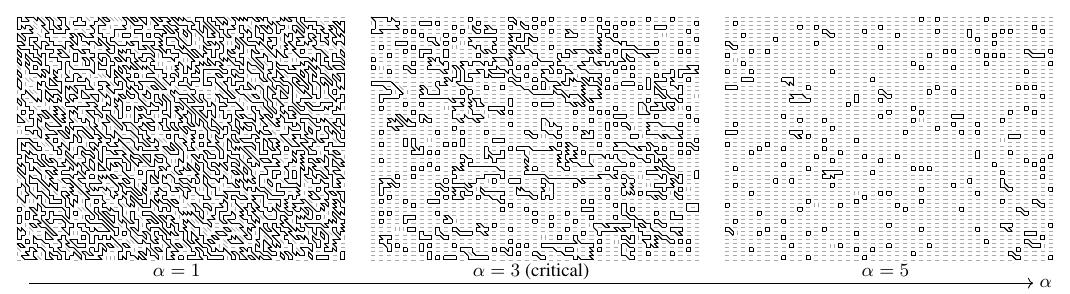}
%\begin{tikzpicture}
%\node[above right] at (0,.1) {
%\reflectbox{\includegraphics[width=.33\textwidth]{double_dimer_a1_n82_75.pdf}}
%\reflectbox{\includegraphics[width=.33\textwidth]{double_dimer_a3_n82_75.pdf}}
%\reflectbox{\includegraphics[width=.33\textwidth]{double_dimer_a5_n82_75.pdf}}
%};
%\draw[->] (0.5,0)--(17.5,0) node[right] {$\wt$};
%\node[above] at (3,0) {$\wt=1$};
%\node[above] at (9,-.05) {$\wt=3$ (critical)};
%\node[above] at (15,0) {$\wt=5$};
%\end{tikzpicture}
\caption{
The quantum dimer model on the weighted triangular lattice as given in Eq.~\eqref{eqn:Htri} exhibits a continuous quantum phase transition as an edge weight parameter $\wt$ varies. Statistical behavior in the different phases and at the critical point can be seen through the behavior of random double-dimer coverings in the corresponding \emph{classical} model, shown here for the $2\times1$ periodic triangular lattice (defined in Fig.~\ref{fig:triangle}) with $\wt=1$ (left, quantum spin liquid), $\wt=3$ (center, critical), and $\wt=5$ (right, columnar order), on a width $81$ grid. Double-edges are shown in gray, while loops are shown in black. In the spin liquid phase $\wt<3$, the double-dimer covering forms many large macroscopic loops, while in the ordered phase $\wt>3$, there are only small loops and double edges.
The ordered phase for $\wt>3$ is a columnar phase with a nonzero fraction of defects.
This loop behavior can be used to intuitively explain the observed vison correlator behavior in the quantum dimer model.
The double-dimer coverings are generated using the Kasteleyn matrix method to compute conditional edge probabilities.
}\label{fig:double}
\end{figure*}

Despite this broad scope, theoretical progress remains hampered by the scarcity of rigorous tools for analyzing these inherently strongly-correlated models. A notable exception is the construction by Rokhsar and Kivelson (RK) \cite{rokhsar1988superconductivity}, which reverse-engineers a local quantum dimer Hamiltonian whose ground state at the RK point is a uniform superposition of dimer coverings, known as the RK wavefunction. 
At such points, equal-time correlators of diagonal operators reduce to those of a corresponding classical dimer model, allowing one to import powerful tools of classical statistical mechanics and integrable probability, including Kasteleyn’s Pfaffian technique on planar graphs \cite{kasteleyn1961statistics,kasteleyn1963dimer,temperley1961dimer,fisher1961stastical} and its subsequent developments. In this work we generalize the RK construction to produce local Hamiltonians whose ground states realize \emph{arbitrary} edge-weighted superpositions of dimer coverings on arbitrary graphs, where a standard RK Hamiltonian can be defined. 
This opens the door to exactly solvable models for a much wider class of spin liquids and ordered phases, as well as controlled examples of transitions between them.
Furthermore, our Hamiltonian construction implies that a vast number of results derived for classical edge-weighted dimer models~\cite{cohn2001variational,kenyon2006dimers,boutillier2007pattern,kenyon2009lectures,cjy2015asymptotic,chhita2014coupling,chhita2016domino,berggren2019correlation,duits2021the,gorin2021lectures,borodin2023biased,berggren2023geometry,boutillier2025focks} are now applicable to a corresponding edge-weighted quantum dimer model.
For example, the frozen corners and multiple so-called ``smooth'' or ``gaseous" regions seen in a $2\times 2$ periodic Aztec diamond will also be present in the corresponding quantum dimer model using our construction. 
We focus here on an analytically tractable quantum dimer model on the triangular lattice that exhibits a continuous quantum phase transition between a topological $\mathbb{Z}_2$ spin liquid and a symmetry-broken ordered state (Fig.~\ref{fig:double}). 

%\tikzexternalenable

\textit{Review of the RK Hamiltonian}--- The quantum dimer Hilbert space is the span of the orthonormal basis $\{|\dconfig\rangle\}$, where $\dconfig$ denotes a dimer configuration.
The RK Hamiltonian \cite{rokhsar1988superconductivity}, written on the square lattice, is
%\tikzexternaldisable
\begin{multline}
    \label{eqn:rk}
    H = -J \sum_{\square} \left[\ket{\plaqA} \bra{\plaqB } + \text{h.c.}\right] \\
    + V \sum_{\square} \left[\ket{\plaqA} \bra{\plaqA} +\ket{\plaqB} \bra{\plaqB}\right]~,
\end{multline}
where the sum is over all plaquettes of the lattice. 
At the RK point, when $J=V$, the exact ground state is solvable and is a uniform superposition of all possible dimer configurations. 
However, in order to witness the rich phenomena arising in weighted classical dimer models, we seek to construct a generalized RK Hamiltonian whose ground state is a weighted superposition of all possible dimer configurations.

\textit{Exact Hamiltonian for weighted dimer coverings}--- In this section, we define an explicit Hamiltonian whose exact ground state can be determined and has dimer coverings appearing with amplitudes proportional to a product of the edge weights.
This construction falls into a class of generalized RK Hamiltonians considered in Refs.~\cite{henley2004classical,castelnovo2005quantum}.
We also consider uniqueness of the ground state, which in general does not follow from the usual identification of an exactly solvable ground state.
For simplicity, we start with the square lattice, but the construction can be generalized to any graph with an RK Hamiltonian.

\begin{figure}[htbp]
\includegraphics{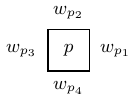}
%  \centering
%  \begin{tikzpicture}[scale=.7]
%    % square
%    \draw[line width=0.7pt] (-0.5,-0.5) rectangle (0.5,0.5);
%    % center label
%    \node at (0,0){$p$};
%    % side labels
%    \node[right=6pt] at (0.3,0){$w_{p_1}$};
%    \node[above=6pt] at (0,0.3){$w_{p_2}$};
%    \node[left=6pt]  at (-0.3,0){$w_{p_3}$};
%    \node[below=6pt] at (0,-0.3){$w_{p_4}$};
%  \end{tikzpicture}
  \caption{A plaquette $p$ with surrounding weights $w_{p_1}$, $w_{p_2}$, $w_{p_3}$, and $w_{p_4}$ in a clockwise order.}
  \label{fig:square-weights}
\end{figure}

Consider the square lattice and denote its set of edges by $\mathcal{E}$.
For each edge $e$, we assign a weight $w_e$, which can be a complex number.
The weight of a dimer covering $W(\mathcal{C})$ is the product of the weights of all the edges that are occupied by dimers in $\mathcal{C}$, 
$W(\mathcal{C}) = \prod_{e \in \mathcal{C}} w_{e}$.
We now construct a Hamiltonian whose ground state is
\begin{equation}
\label{eqn:weighted-RK-wavefunction}
    \ket{\psi_{\mathbf{w}}} = \sum_{\mathcal{C}} W(\mathcal{C}) \ket{\mathcal{C}}~,
\end{equation}
where $\mathbf{w}$ is a list of weights $w_e$.
Note that the probability of $\mathcal{C}$ in \cref{eqn:weighted-RK-wavefunction} is $|W(\mathcal{C})|^2$, rather than $|W(\mathcal{C})|$ which is the standard convention in edge-weighted classical dimer model literature.
For convenience, we use the former convention for the general construction.
The edge-weighted RK Hamiltonian having  $\ket{\psi_{\mathbf{w}}}$ as a ground state when $J=V$ is given by:
\begin{multline}
    \label{eqn:weighted-hamiltonian}
    H = -J \sum_{\square} w_{p_2}^* w_{p_4}^* w_{p_1} w_{p_3} \ket{\plaqA} \bra{\plaqB } + \text{h.c.} \\
    + V \sum_{\square} |w_{p_2} w_{p_4} |^2 \ket{\plaqA} \bra{\plaqA} + |w_{p_1} w_{p_3}|^2 \ket{\plaqB} \bra{\plaqB}~,
\end{multline}
%\tikzexternalenable
where the sum is over all the square plaquettes $p$, with $p_i$ for $i\in\{1,2,3,4\}$ denoting the four edges of $p$ taken in a clockwise order (see~\cref{fig:square-weights}).
For $J=V$, analogous to the RK point of Eq.~\eqref{eqn:rk}, the Hamiltonian turns into a sum of (noncommuting) projectors
    \(H = J \sum_{\square} P_{\square}~,\)
where
%\tikzexternaldisable
\begin{equation}
    P_{\square} \!= \!\left[ w_{p_2}^*w_{p_4}^* \ket{\plaqA}\! -\! w_{p_1}^*\!w_{p_3}^* \ket{\plaqB}  \right]\!\! \left[ w_{p_2} \!w_{p_4} \bra{\plaqA}\! -\! w_{p_1} \!w_{p_3} \bra{\plaqB}  \right].
\end{equation}
It is easy to see that $P_{\square}^2 = \left[ |w_{p_2} w_{p_4}|^2 + |w_{p_1} w_{p_3}|^2\right] P_{\square}$, implying $P_{\square}$ is a projector up to a positive scalar factor.
This immediately implies that all the eigenenergies must be nonnegative.
Moreover, it can be explicitly checked that $P_{\square} \ket{\psi_{\mathbf{w}}} = 0$, for all $P_{\square}$.
Thus $\ket{\psi_{\mathbf{w}}}$ is a ground state of $H$.
Note that this construction is very general and applies to any configuration of weights, which need not be translationally invariant. 
It is worth noting that $\ket{\psi_{\mathbf{w}}}$ is not necessarily the unique ground state for an arbitrary graph; however on any simply connected finite region of the square lattice, it is the unique ground state of $H$, as long as none of the weights are zero, since any two dimer coverings on such a region are connected by ring flips \cite{thurston1990conways}. 
Any diagonal observable calculated in $\ket{\psi_{\mathbf{w}}}$ will be equal to the observable calculated in a classical dimer model with covering weights proportional to $\abs{W(\mathcal{C})}^2$.
We can thus use results and techniques for weighted classical dimer models to draw conclusions about the quantum dimer model defined above.

\begin{figure}[htb]
\includegraphics{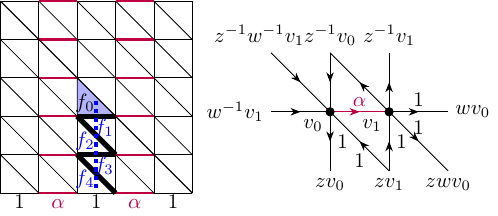}
\caption{
(Left) Path in the triangular lattice for the dimer-dimer and vison correlator calculations. The dashed blue vertical path $\gamma$ is used for calculating the vison correlator between faces $f_0$ and $f_\ell$.
(Right) Fundamental domain and Kasteleyn orientation for the $2\times1$ periodic triangular lattice with one horizontal edge weight $\wt$ and all five other weights equal to 1. 
Horizontal and vertical translates of the fundamental domain are indexed by variables $w,z\in\C$. 
}\label{fig:triangle}
\end{figure}

\textit{Weighted dimer Hamiltonian for the triangular lattice}--- Similar to the square lattice Hamiltonian, we can write a Hamiltonian on the triangular lattice. 
The only modification is that we sum over several kinds of plaquettes. In order to ensure uniqueness of the ground state using ergodicity with $4$ and $6$-cycles \cite{hartarsky2024local,kenyon1996perfect}, we include larger $6$-cycle plaquettes as well.
The edge weights we consider on the lattice are $2\times1$ periodic, i.e., they are $2$-periodic in the horizontal direction and $1$-periodic in the vertical direction, allowing for 6 distinct edge weights. 
For simplicity, we shall take one horizontal edge weight to be $\wt\ge0$, and all five other edge weights to be 1 (see Fig.~\ref{fig:triangle}). 
For these weights, the Hamiltonian using the above construction is:
\begin{multline}\label{eqn:Htri}
    H = \sum_{\squareone}
    \mathcal{P} \left( \ket{\plaqA} - \ket{\plaqB}\right) + 
    \sum_{\squarewt}
    \mathcal{P} \left( \wt \ket{\plaqA} - \ket{\plaqB}\right) \\
    + \sum_{\parallelograma} \mathcal{P} \left(\sqrt{\wt} \ket{\parallelogramaA} - \ket{\parallelogramaB} \right) + \sum_{\parallelogramb} \mathcal{P} \left( \Big|{\parallelogrambA}\Big\rangle - \Big|{\parallelogrambB}\Big\rangle \right)\\
    +\text{\{6-cycle terms\}},
    % + \!\!\sum_{\triangleAwtone}\!\! \mathcal{P}\Big(\Big|\triangleAone\Big\rangle \!-\! \sqrt{\wt}\Big|\triangleAtwo\Big\rangle\Big)\!+\!
    % \sum_{\triangleAonewt}\!\! \mathcal{P}\Big(\sqrt{\wt}\Big|\triangleAone\Big\rangle \!-\! \Big|\triangleAtwo\Big\rangle\Big)\\
    % +\!\sum_{\triangleBwtone}\!\! \mathcal{P}\Big(\Big|\triangleBone\Big\rangle \!-\! \sqrt{\wt}\Big|\triangleBtwo\Big\rangle\Big)+\!
    % \sum_{\triangleBonewt} \!\!\mathcal{P}\Big(\sqrt{\wt}\Big|\triangleBone\Big\rangle \!-\! \Big|\triangleBtwo\Big\rangle\Big),
\end{multline}
%\tikzexternalenable
where \{6-cycle terms\} represents a similar sum over projectors of triangular 6-cycle plaquettes, which we write out in full in the Supplemental Material \cite{sm}.
As before, we use the notation $\mathcal{P}(\ket{\psi}) \equiv \ketbra{\psi}$, and the subscripts in the summations indicate the types of plaquettes in each sum. 
The edge weight $\sqrt{\wt}$ is used in the Hamiltonian so that correlations in the ground state are given by correlations of the classical model with weight $\wt$. 
Since Ref.~\cite{hartarsky2024local} showed that any two dimer coverings of a rectangular domain in the triangular lattice can be connected by $4$ and $6$-cycle flips, $\ket{\psi_{\mathbf{w}}}$ in Eq.~\eqref{eqn:weighted-RK-wavefunction} is the \emph{unique} ground state of the above Hamiltonian on such domains. 
(See Supplemental Material \cite{sm} for further details.)
Note that uniqueness does not hold on the torus, where the Hilbert space splits into different topological sectors \cite{moessner2001resonating,moessner2011introduction}.

\textit{Topological phase transition on the triangular lattice}--- In contrast to quantum dimer models considered on bipartite lattices
\cite{henley1997relaxation,fradkin2004bipartite,moessner2011introduction},
the RK quantum dimer model on the unweighted triangular lattice is known to be in a $\Z_2$ quantum spin liquid phase~\cite{moessner2001resonating,misguich2008quantum,ioselevich2002groundstate,moessner2011introduction}. 
By considering non-uniform edge weights on the triangular lattice, we show that the quantum dimer model defined in Eq.~\eqref{eqn:Htri} undergoes a topological phase transition from a $\Z_2$ quantum spin liquid to a columnar ordered phase.

The $2\times1$ periodic weights on the triangular lattice were first introduced and studied for the \emph{classical} dimer model in Ref.~\cite{nash2017dimer}, where the authors studied the amoebas and geometry of the magnetically-altered Kasteleyn matrix \cite{kenyon2006dimers}, showing an associated Chern number change indicates a classical phase transition at $\wt=3$, although the nature of the phases and type of phase transition are not explored.
In this Letter, we consider the \emph{quantum} dimer model on the $2\times1$ periodic triangular lattice, and show that the model undergoes a quantum phase transition, characterized by the \emph{vison} correlator and the topological min-entropy, across $\wt=3$.
We note by construction the quantum dimer model of Eq.~\eqref{eqn:Htri} is exactly solvable, including through the transition point $\wt=3$: The exact ground state is given by Eq.~\eqref{eqn:weighted-RK-wavefunction}, and all correlators can be expressed exactly in terms of the associated Kasteleyn matrix \cite{kasteleyn1961statistics,fisher1961stastical,temperley1961dimer,kasteleyn1963dimer} (a specific weighted adjacency matrix; see Supplemental Material \cite{sm}) and its inverse \cite{temperley1961dimer,fisher1961stastical,montroll1963correlations,kenyon1997local,fendley2002classical}.
In particular, the dimer-dimer and vison correlators defined below, which identify the quantum phase, have exact expressions which can be studied analytically as well as evaluated numerically to high (even arbitrary \cite{wilson1997determinant,lee1995fraction,nakos1997fraction,shah2025breakdown}) precision.
Additionally, as we will see, the phase transition can be detected analytically by the topological min-entropy, which diagnoses topological order, and which changes from the topological value $\log2$ for $\wt<3$, to $0$ for $\wt>3$.

The dimer-dimer correlator is $C_{\mu\nu}^d=\langle d_\mu d_\nu\rangle-\langle d_\mu\rangle\langle d_\nu\rangle$,
where $\mu$ and $\nu$ label edges of the graph. 
Writing $\mu=(\mu_0,\mu_1)$ and $\nu=(\nu_0,\nu_1)$ in terms of the corresponding vertices, $C_{\mu\nu}^d$ can be expressed in terms of the Kasteleyn matrix $K$ as \cite{kenyon1997local,fendley2002classical,cimasoni2007dimers}
\begin{align}\label{eqn:dimerK}
C_{\mu\nu}^d&=K_{\mu_0,\mu_1}K_{\nu_0,\nu_1}\big(\!-\!K^{-1}_{\mu_0\nu_0}K^{-1}_{\mu_1\nu_1}+K^{-1}_{\mu_0\nu_1}K^{-1}_{\mu_1\nu_0}\big).
\end{align}
Similar to the bipartite planar case \cite{cohn2001variational,kenyon2006dimers}, we show in the Supplemental Material \cite{sm} that the entries of $K^{-1}$ in the infinite size periodic limit are given by double integrals of the form
\begin{align}\label{eqn:Kinvgen}
\langle z_0|K^{-1}|z_1\rangle&=\int_0^{2\pi}\int_0^{2\pi}\frac{q_{z_0,z_1}(e^{ik_x},e^{ik_y})}{P(e^{ik_x},e^{ik_y})}\frac{dk_x}{2\pi}\frac{dk_y}{2\pi},
\end{align}
where $q_{z_0,z_1}$ is a polynomial depending on the points $z_0,z_1$, and $P(e^{ik_x},e^{ik_y})$ is the characteristic polynomial, equivalently the product of band dispersions in the free-fermion picture, for the lattice.
This leads to an exact asymptotic expression in terms of sums and products of explicit integrals for any correlator, which allows for analytic study of correlators \cite{cohn2001variational,kenyon2006dimers,basor2007asymptotics,basor2017exact,shah2025breakdown}. In particular, for this model we analytically show \cite{sm} that the dimer-dimer correlations decay exponentially off of the critical point at $\wt=3$. More generally, in the Supplemental Material \cite{sm}, we consider the full six parameter $2\times 1$ edge weights case, and show the dimer-dimer correlator decays exponentially away from the set where one of the horizontal or diagonal edge weights is equal to the sum of the other three remaining such weights.

\begin{figure*}[tb]
\includegraphics[height=1.57in]{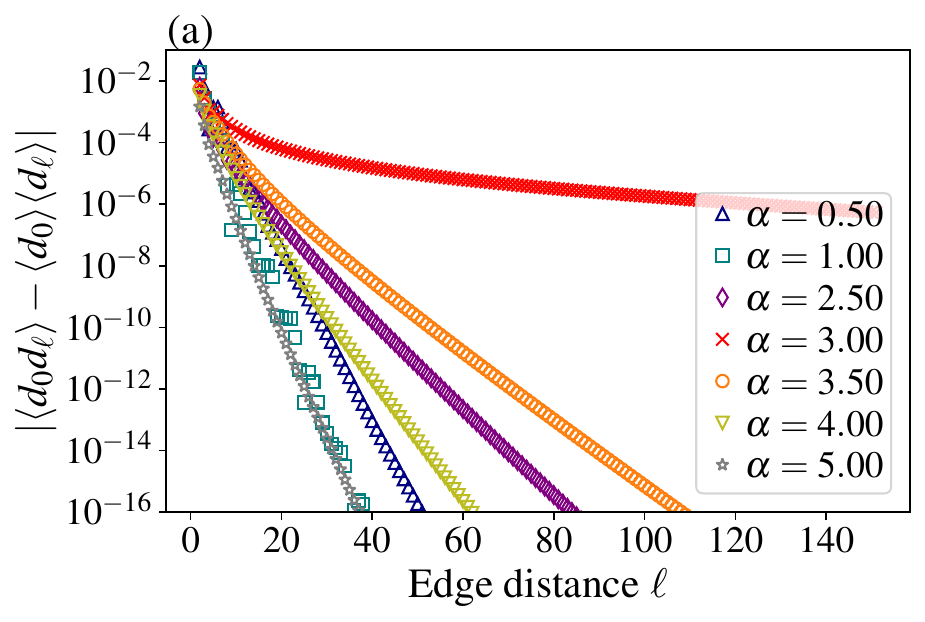}
\includegraphics[height=1.57in]{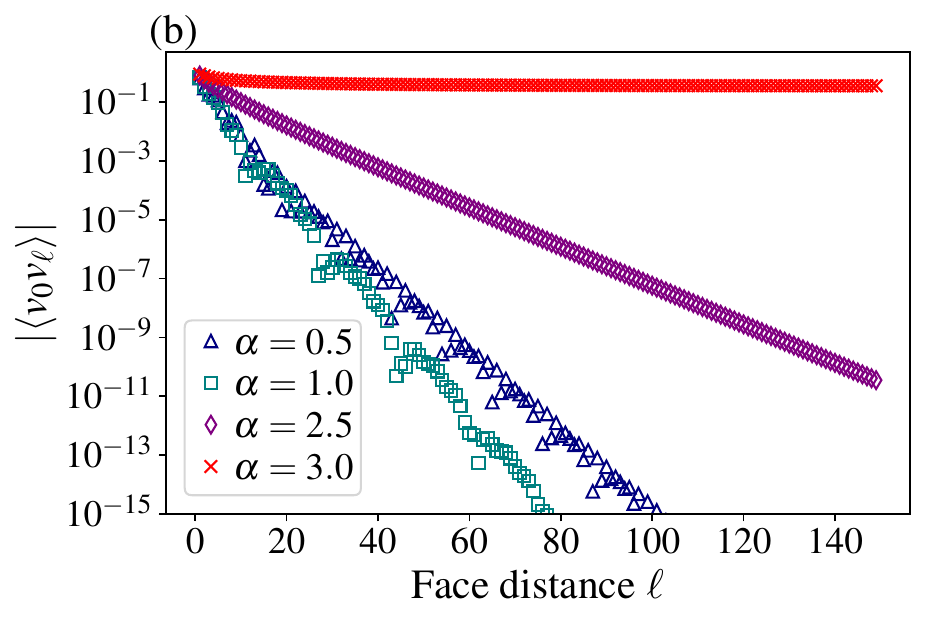}
\includegraphics[height=1.57in]{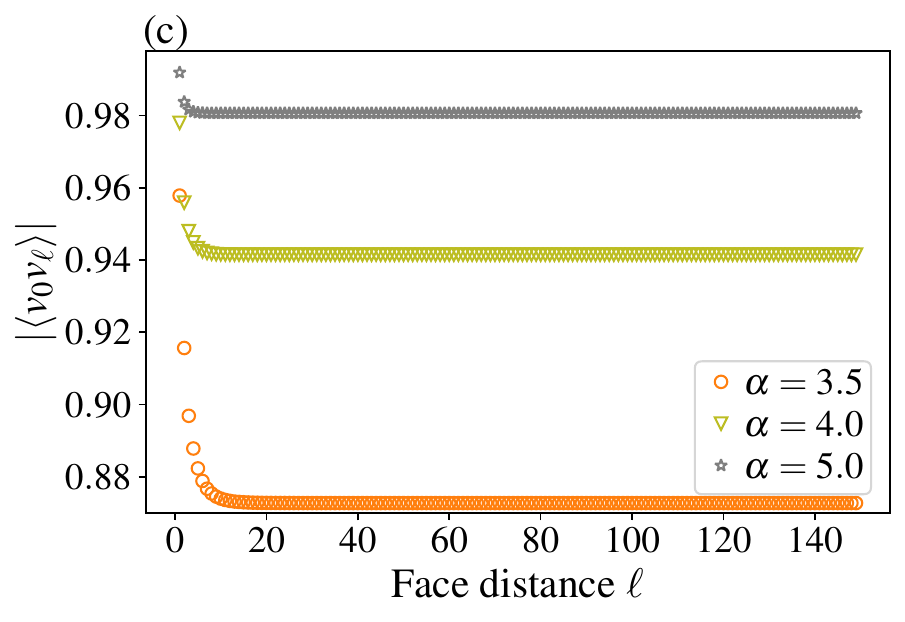}
\caption{
Dimer-dimer and vison correlators for different values of $\wt$, calculated along the path shown in Fig.~\ref{fig:triangle} on a $303\times303$ size grid. The path starts in the center and ends roughly midway to the boundary.
(a)  The dimer-dimer correlator exhibits exponential decay for all values of $\wt\ne3$, which we show analytically in the Supplemental Material \cite{sm}. (b) The vison correlator $|\langle v_0v_\ell\rangle|$ decays exponentially for $\wt<3$, signaling a spin liquid phase. It exhibits power-law decay at the critical value $\wt=3$. 
(c) The vison correlator is constant for $\wt>3$, identifying a trivial ordered phase. 
In all cases, correlators for the adjacent vertical path (passing through weight $\wt$ edges) and a horizontal path demonstrate similar behavior, with some periodic differences mostly depending on the parity of the path distance. We provide these additional plots in the Supplemental Material \cite{sm}.
}\label{fig:visons}
\end{figure*}

Quantum dimer models can be mapped to gauge theories, in particular the model studied here can be mapped to a $\mathbb{Z}_2$ gauge theory~\cite{moessner2001shortranged}. 
Monomers (sites without any dimers touching it) and visons correspond to the electric and magnetic charges under such mapping.
Exponential decay of all dimer-dimer correlators and vison correlators implies that the system is gapped and is a $\mathbb{Z}_2$ quantum spin liquid \cite{senthil2000z2gauge,senthil2001fractionalizationPRB,senthil2001fractionalizationPRL,balents2002fractionalization}.
The vison correlator, which is a nonlocal string operator, can be written in terms of $K$ and $K^{-1}$ as follows.  
For a fixed path $\gamma$ between faces $f_0$ and $f_\ell$ (Fig.~\ref{fig:triangle}), consider the edges $E$ crossed by $\gamma$, and count the number of dimers present in $E$.
The vison correlator between $f_0$ and $f_\ell$ is $|\langle(-1)^{\#\{\text{dimers in $E$}\}}\rangle|$, and can be written in terms of the Kasteleyn matrix as \cite{shah2025breakdown}
\begin{align}\label{eqn:visonK}
\big|\langle(-1)^{\#\{\text{dimers in $E$}\}}\rangle\big|&=\sqrt{\det\left(I_{2|E|}-2(K')_E(K^{-1})_{E}\right)},
\end{align}
where $I_{2|E|}$ is the $2|E|\times2|E|$ identity matrix, $(K^{-1})_E$ is the restriction of $K^{-1}$ to $E$ (i.e. to the $2|E|$ vertex indices associated with $E$) and $(K')_E$ is a restricted Kasteleyn matrix with nonzero entries only for edges in $E$.

\begin{figure}[htb]
\includegraphics[width=.37\textwidth]{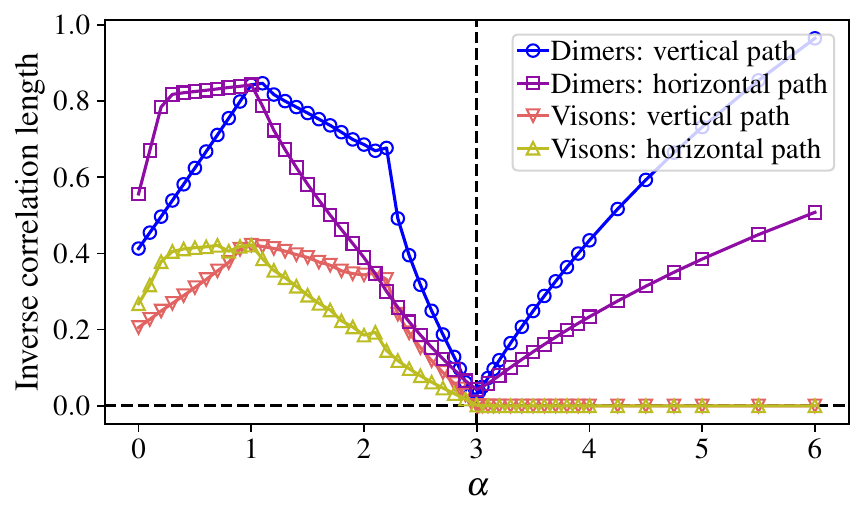}
\caption{Inverse correlation lengths for the dimer-dimer and vison correlators across a vertical and horizontal path for system size $303\times303$, showing linear behavior near $\wt=3$. 
The inverse correlation length is calculated with respect to the edge or face distance, which is half the coordinate or lattice. 
The labeled vertical path goes through horizontal edges with weight 1, but the values for the adjacent path through edges with weight $\wt$ behave similarly.
}\label{fig:corr}
\end{figure}

\begin{figure}[htb]
\includegraphics[width=0.37\textwidth]{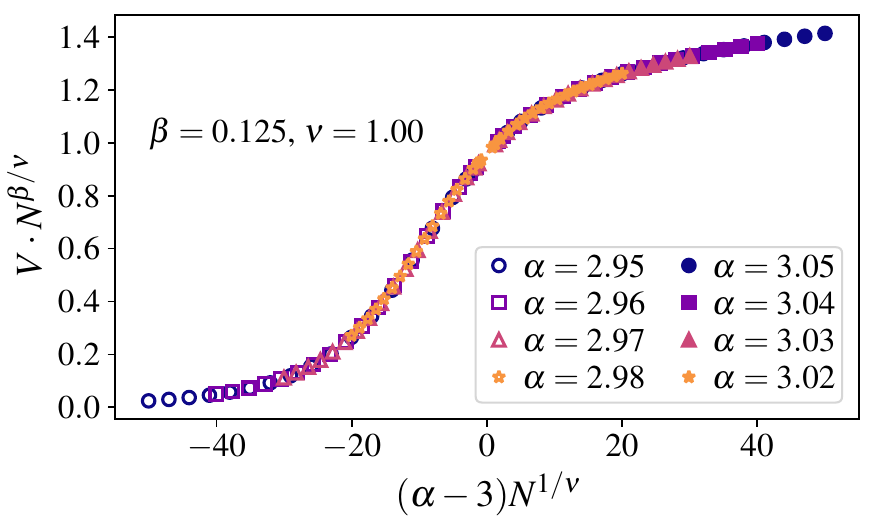}
\caption{Finite-size scaling of the vison order parameter.
In the thermodynamic limit, the long-distance vison order parameter is   $V_\infty=\lim_{\ell\to\infty}\sqrt{|\langle v_0v_\ell\rangle|}$, where $|\langle v_0v_\ell\rangle|$ is the vison correlator. 
However, on a finite $N\times N$ lattice, we use the finite-size estimator $V=\sqrt{|\langle v_0v_\ell\rangle|}$ evaluated at lattice distance $\ell\simeq N/4$, halfway from the center to the edge.
The path is the vertical path through weight $\wt$ horizontal edges starting in the center, and calculated for system sizes $N=41,61$, and $101$ through $1001$ by $60$s.
The curve collapse occurs for $\beta=1/8$ and $\nu=1$ as shown.
Similar curve collapse for other paths are shown in the Supplemental Material \cite{sm}.
}\label{fig:fs_scaling}
\end{figure}

Having defined the relevant correlators and discussed their exact expressions, we plot them  in Fig.~\ref{fig:visons}, and summarize the quantum phases of our model as a function of the parameter $\wt$ as follows.
For $\wt<3$, the dimer-dimer and vison correlators decay exponentially, indicating a $\Z_2$ quantum spin liquid phase. For $\wt=3$, the correlators are critical. For $\wt>3$, the dimer-dimer correlator decays exponentially, but the vison correlator is constant, indicating a trivial ordered phase. 
The analytic derivations of the exponential decay of the dimer-dimer correlator as well as the analytic estimates of the real space Green's function correlation lengths are given in the Supplemental Material \cite{sm}.
The diverging correlation lengths for the dimer-dimer and vison correlators near the transition point are shown in Fig.~\ref{fig:corr}. 
While it appears there may be cusps at values of $\wt\ne3$, it is possible this may be a numerical artifact; importantly, our topological diagnostic does not show any transition within the $\alpha<3$ region.
The critical exponents $\beta=1/8$ and $\nu=1$ for the transition, which are consistent with the 2D Ising universality class, are obtained via finite-size scaling numerics up to lattice size $1001\times1001$ in Fig.~\ref{fig:fs_scaling}.
Additionally, we consider the topological R\'enyi entropy of order $\infty$ (topological min-entropy) $s_\infty$, which along with the topological entanglement entropy (TEE) and other topological R\'enyi entropies, is used to detect topological order in many-body systems \cite{kitaev2006topological,levin2006detecting,flammia2009topological}. 
In the setting of quantum dimer models, the von Neumann and R\'enyi entropies at the RK point can be expressed in terms of classical boundary probabilities \cite{furukawa2007topological,papanikolaou2007topological,stephan2012renyi,selem2013entanglement}. In the Supplemental Material \cite{sm}, we use the method of Ref.~\cite{stephan2012renyi} to analytically show
\begin{align}\label{eqn:s-inf}
s_\infty=\log2\text{ for }\wt<3,\;\text{ and }=0\text{ for }\wt>3.
\end{align}
This analytically confirms the topological nature of the phase transition at $\wt=3$, from the topological $\Z_2$ spin liquid phase to a topologically trivial phase.

All numerical correlator values in the plots are calculated using the Kasteleyn matrix method (with standard floating point sparse LU methods to calculate entries of $K^{-1}$), using exact expressions Eqs.~\eqref{eqn:dimerK} and \eqref{eqn:visonK} for the correlators. 
We emphasize that while we present some finite-size numerical results in this section, by Eq.~\eqref{eqn:Kinvgen} we obtain exact formulas for all correlators in the infinite-size limit, which are given in terms of products and sums of explicit integrals. For example, the real space Green's function or entries of $K^{-1}$ between two vertically separated points $0$ and $y$ in the infinite-size limit is given as: 
\begin{multline}\label{eqn:Kinv-int2}
\langle 0|K^{-1}|y\rangle=\\
\frac{\pm i}{4\pi^2}\int_0^{2\pi}\!\!\!\int_0^{2\pi}\!\frac{2\sin(k_y)e^{-ik_yy}\,dk_x\,dk_y}{4\sin^2k_y+|\wt-e^{ik_y}-e^{-ik_x}-e^{-ik_x}e^{-ik_y}|^2},
\end{multline}
with the sign depending on the column parity. The integral can also be turned into a single integral using contour integration.
Sums and products of such integrals determine all correlations in the infinite-size limit, including dimer-dimer and vison correlators as seen by Eqs.~\eqref{eqn:dimerK} and \eqref{eqn:visonK}.

Using this, we analytically obtain exponential decay of dimer-dimer correlators off of the critical point, equivalent to the presence of a band gap in the fermion model.
Additionally, we can analytically identify the divergence of the correlation length for $K^{-1}$, equivalently for the real-space Green's function in the free-fermion picture, as $\wt\to3$.
As detailed in the Supplemental Material \cite{sm}, we analytically obtain the Green's function correlation length along a vertical path as
$\xi_\mathrm{G}\propto{1}/{|\wt-3|}$, as $\wt\to3$.
The numerical results in Fig.~\ref{fig:corr} also show that as $\wt\to3$,
$\xi_\mathrm{dimer}\propto|\wt-3|^{-1}$ and $\xi_\mathrm{vison}\propto(3-\wt)_+^{-1}$,
where $(x)_+=\max(0,x)$.
For $\wt>3$, the vison correlator in Fig.~\ref{fig:visons}(c) is constant so the correlation length is $\infty$.
Regarding the vison correlator, we note that it may be possible to analytically investigate the constant vison correlator values using the methods of Ref.~\cite{basor2007asymptotics}. 
A simpler vison correlator constant was evaluated for the (uniform) RK dimer model on the square-octagon lattice in Ref.~\cite{shah2025breakdown}.
Finally, we mention that the expressions for the infinite-size Green's function limit are also used to analytically show the topological min-entropy transition in Eq.~\eqref{eqn:s-inf}.

To understand the nature of the phases and explain the constant vison correlator, we consider double-dimer coverings (see Fig.~\ref{fig:double}); that is, the union $C\cup C'$ of two independent random dimer coverings $C,C'$. 
Since each vertex will have two edges present, the double-dimer covering consists only of loops and double edges. 
For critical bipartite lattices, the double-dimer model is expected to behave as the Conformal Loop Ensemble $\mathrm{CLE}_4$ \cite{kenyon2014conformal,dubedat2019double,basok2021tau}.
Here we are primarily interested in understanding the vison correlator in the non-critical spin liquid or ordered phases. 
The vison correlator can be expressed in terms of the double-dimer coverings $C\cup C'$ as
\begin{align}
|\langle v_0v_\ell\rangle|^2 &= \langle(-1)^{\#\text{ dimers in $C\cup C'$ along $\gamma$}}\rangle,
\end{align}
where $\gamma$ is a path between the two faces.
Since crossing a double edge does not change the dimer parity, 
the vison correlator can be written in terms of the number of loops in $C\cup C'$ surrounding each endpoint of $\gamma$.
As seen in Fig.~\ref{fig:double}, in the quantum spin liquid phase, loops are large and complicated, allowing for many changes in parity along a vison path. 
In contrast, in the trivial ordered phase, loops are small and not very common, appearing similar to numerical simulations of the smooth or gaseous phase of bipartite periodic lattices \cite{kenyon2006dimers}. This sparse loop behavior can be argued to produce a constant vison correlator, as explained further in the Supplemental Material \cite{sm}.

\textit{Conclusion}---
In this Letter, we generalized the quantum dimer construction of Rokhsar and Kivelson \cite{rokhsar1988superconductivity} to construct quantum dimer Hamiltonians whose ground state at the RK point is an arbitrary edge-weighted superposition of dimer coverings. Using this construction, we demonstrated the existence of an exactly solvable topological phase transition in the weighted quantum dimer model on the triangular lattice, from a $\Z_2$ quantum spin liquid phase to a columnar ordered phase. We used both analytical and numerical results to identify the quantum phases, using the real space Green's function, dimer-dimer correlator, vison correlator, and topological min-entropy.
Furthermore, we identified critical exponents $\beta=1/8$ and $\nu=1$, which are consistent with those of the 2D Ising universality class.

This work opens several new directions for future research on quantum phase transitions, topological phases, and statistics of dimer models. A natural question is to study the model away from the RK point, and to investigate the behavior and universality class of the phase transition there. Another direction is to develop a general theory and classification for non-bipartite dimer models, which does not yet exist in comparison to the theory for bipartite models \cite{kenyon2006dimers}. 
Further study of the behavior of weighted quantum dimer models at the exactly solvable RK point is also a promising direction, in particular it would be interesting to identify all possible phases and topological orders that one can obtain with such models, for example in the direction of double semion phases investigated in Refs.~\cite{qi2015double,buerschapper2014double}, or using the full six parameters in the $2\times1$ model (or more general models).

% prevent acknowledgments and bibliography from showing up in Supplemental Material table of contents
% \let\oldaddcontentsline\addcontentsline
% \renewcommand{\addcontentsline}[3]{}

\begin{acknowledgments}
\textit{Acknowledgments}---J.S. and V.G. were supported by the US Army Research Office under Grant Number W911NF-23- 1-024, Schwinger Foundation, and Simons Foundation. The work of Amol Aggarwal was partially supported by a Clay Research Fellowship and a Packard Fellowship for Science and Engineering. Alexei Borodin was supported by NSF grant DMS-2450323 and the Simons Investigator program.
\end{acknowledgments}

\textit{Data availability}---The code used is available~\cite{shoudimerstriangular}.

% supplement refs
\nocite{hartarsky2024local,kenyon1996perfect,kasteleyn1961statistics,kenyon2009lectures,cimasoni2014geometry,gorin2021lectures,kasteleyn1967graph,fendley2002classical,kenyon2014conformal,cimasoni2007dimers,kenyon2006dimers,shah2025breakdown,cohn2001variational,broer2011dynamical,arnold1987geometrical,fisher1963statistical,kenyon2008height,boutillier2023minimal,henley1997relaxation,fradkin2004bipartite,moessner2011introduction,kitaev2006topological,levin2006detecting,furukawa2007topological,papanikolaou2007topological,stephan2012renyi,selem2013entanglement,flammia2009topological}

\bibliography{dimers}

\end{document}

\title{Exactly Solvable Topological Phase Transition in a Quantum Dimer Model\\ Supplemental Material}

\author{Laura Shou\,\orcidlink{0000-0001-8624-8063}}
\affiliation{Joint Quantum Institute, Department of Physics, University of Maryland, College Park, MD 20742, USA}
\author{Jeet Shah\,\orcidlink{0000-0001-5873-8129}}
\affiliation{Joint Quantum Institute, Department of Physics, University of Maryland, College Park, MD 20742, USA}
\affiliation{Joint Center for Quantum Information and Computer Science, NIST/University of Maryland,
College Park, MD 20742, USA}
\author{Matthew Lerner-Brecher}
\affiliation{Department of Mathematics, Massachusetts Institute of Technology, Cambridge, MA 02139, USA}
\author{Amol Aggarwal\,\orcidlink{0000-0002-8091-2215}}
\affiliation{Department of Mathematics, Stanford University, Stanford, CA 94305, USA}
\author{Alexei Borodin\,\orcidlink{0000-0002-2913-5238}}
\affiliation{Department of Mathematics, Massachusetts Institute of Technology, Cambridge, MA 02139, USA}
\author{Victor Galitski}
\affiliation{Joint Quantum Institute, Department of Physics, University of Maryland, College Park, MD 20742, USA}

%\tikzexternaldisable
\maketitle
%\tikzexternalenable
In the Supplemental Material, we provide additional details, proofs, and analytical and numerical results to support the Main Text.

\tableofcontents

\section{Weighted Hamiltonian details}

In this section, we provide further details for construction and ground state uniqueness of the Hamiltonian in Eq.~\mainhamiltonian\ of the Main Text.
Previous quantum dimer Hamiltonians on the triangular lattice typically include only unweighted 4-cycle ring flip terms.
However, as observed in \cite{hartarsky2024local}, 4-cycle ring flips may not always be sufficient, even in planar rectangular domains, to connect two dimer coverings on the triangular lattice. While demonstrated to be insufficient for a small rectangular domain, it remains an open question whether 4-cycle ring flips are enough for sufficiently large rectangular domains \cite{hartarsky2024local}.
Ref.~\cite{kenyon1996perfect} showed that one can use 4, 6, and 8-cycle flips to connect any two dimer coverings on a simply connected region of the triangular lattice, and Ref.~\cite{hartarsky2024local} reduced this to only 4 and 6-cycle flips for rectangular domains. 
We use this latter result to obtain ground state uniqueness for the Hamiltonian defined in Eq.~\mainhamiltonian\ of the Main Text.

First we give the general construction for the weighted projectors in the Hamiltonian for an arbitrary plaquette, although we will only need $4$ and $6$-cycle plaquettes.
Let $A$ be a plaquette whose boundary consists of $2\ell$ edges $e_1,\ldots,e_{2\ell}$ traversed in order, with corresponding edge weights $w_1,\ldots,w_{2\ell}$. 
Let $\ket{135\ldots(2\ell-1)}$ denote the dimer configurations with dimers on edges $e_1,e_3,\ldots,e_{2\ell-1}$ of $A$, and $\ket{246\ldots(2\ell)}$ those with dimers on edges $e_2,e_4,\ldots,e_{2\ell}$. 
Then using the notation $\mathcal{P}(\ket{\psi}) \equiv \ketbra{\psi}$, we define analogously to Eq.~\mainproj\ of the Main Text,
\begin{align}
P_A:=\mathcal{P}\left\{(w_2w_4\cdots w_{2\ell})^*\ket{135\ldots(2\ell-1)}-(w_1w_3\ldots w_{2\ell-1})^*\ket{246\ldots(2\ell)}\right\}.
\end{align}
More explicitly, let $V_{A_1}$ be the projection onto the set of dimer configurations such that $A$ is flippable and has dimers on edges $e_1,\ldots,e_{2\ell-1}$, and let $V_{A_2}$ be the projection onto the set of dimer configurations such that $A$ is flippable and has dimers on edges $e_2,\ldots,e_{2\ell}$.
Let $F_{A_{12}}$ be the operator which flips the plaquette $A$ from dimers on $e_1,e_3,\ldots,e_{2\ell-1}$ to dimers on $e_2,e_4,\ldots,e_{2\ell}$, and zero otherwise, and let $F_{A_{21}}$ be its hermitian conjugate.
Then letting $v_1:=w_1w_3\ldots w_{2\ell-1}$ and $v_2:=w_2w_4\cdots w_{2\ell}$, one can expand $P_A$ as
\begin{align}
P_A&=|v_2|^2V_{A_1}+|v_1|^2V_{A_2}-(v_2^*v_1 F_{A_{12}}+\mathrm{h.c.}).
\end{align}
In either form, one can explicitly calculate that $P_A^2=(|v_1|^2+|v_2|^2)P_A$, which implies $P_A\ge0$.

\begin{figure}[htb]
\includegraphics{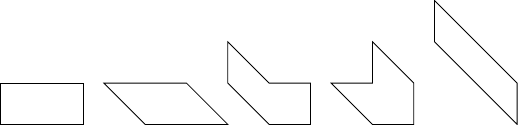}
%\begin{tikzpicture}[scale=.7]
%\draw(0,0)--(2,0)--(2,1)--(0,1)--cycle;
%\draw[xshift=2.5cm] (1,0)--(3,0)--(2,1)--(0,1)--cycle;
%\draw[xshift=5.5cm] (0,1)--(1,0)--(2,0)--(2,1)--(1,1)--(0,2)--cycle;
%\draw[xshift=8cm] (0,1)--(1,0)--(2,0)--(2,1)--(1,2)--(1,1)--cycle;
%\draw[xshift=10.5cm,yshift=1cm] (0,1)--(2,-1)--(2,0)--(0,2)--cycle;
%%\draw[xshift=13cm] (0,2)--(1,2)--(2,1)--(2,0)--(1,0)--(0,1)--cycle; % not a flippable plaquette
%\end{tikzpicture}
\caption{Examples of alternating 6-cycle plaquettes in the triangular lattice which are flippable using 4-cycle ring flips, and thus do not need to be included in the Hamiltonian \eqref{eqn:Htri2}.
}\label{fig:6cycles}
\end{figure}

The full Hamiltonian from Eq.~\mainhamiltonian\ of the Main Text, written with the 6-cycle terms, can be taken to be
%\tikzexternaldisable
\begin{multline}\label{eqn:Htri2}
    H = \sum_{\squareone}
    \mathcal{P} \left( \ket{\plaqA} - \ket{\plaqB}\right) + 
    \sum_{\squarewt}
    \mathcal{P} \left( \wt \ket{\plaqA} - \ket{\plaqB}\right) 
    + \sum_{\parallelograma} \mathcal{P} \left(\sqrt{\wt} \ket{\parallelogramaA} - \ket{\parallelogramaB} \right) + \sum_{\parallelogramb} \mathcal{P} \left( \Big|{\parallelogrambA}\Big\rangle - \Big|{\parallelogrambB}\Big\rangle \right)\\
    + \sum_{\triangleAwtone}\mathcal{P}\Big(\Big|\triangleAone\Big\rangle -\sqrt{\wt}\Big|\triangleAtwo\Big\rangle\Big)+
    \sum_{\triangleAonewt}\mathcal{P}\Big(\sqrt{\wt}\Big|\triangleAone\Big\rangle -\Big|\triangleAtwo\Big\rangle\Big)
    +\!\sum_{\triangleBwtone}\mathcal{P}\Big(\Big|\triangleBone\Big\rangle - \sqrt{\wt}\Big|\triangleBtwo\Big\rangle\Big)+
    \sum_{\triangleBonewt} \mathcal{P}\Big(\sqrt{\wt}\Big|\triangleBone\Big\rangle - \Big|\triangleBtwo\Big\rangle\Big),
\end{multline}
%\tikzexternalenable
where as before, we use the notation $\mathcal{P}(\ket{\psi}) \equiv \ketbra{\psi}$, and use the subscripts in the summations to indicate the types of plaquettes in each sum. 
The first row of \eqref{eqn:Htri2} is the sum over all types of 4-cycle plaquettes, while the second row is a sum over 6-cycle plaquettes. There are other alternating 6-cycle plaquettes (Fig.~\ref{fig:6cycles}) not included in \eqref{eqn:Htri2}, but they can be flipped with 4-cycle ring flips so do not need to appear in the Hamiltonian.

The Hamiltonian \eqref{eqn:Htri2} on a rectangular domain in the triangular lattice has the unique ground state $|\psi_{\mathbf{w}}\rangle$ given by Eq.~\maings\ of the Main Text, with the probability of $|\mathcal C\rangle$ given by the classical weighted probability of the configuration $\mathcal C$.
To see the above, first note that $H\ge0$ since $P_A\ge0$ for every projection term in \eqref{eqn:Htri2}, and that $H|\psi_{\mathbf w}\rangle=0$ since one can check $P_A|\psi_{\mathbf w}\rangle=0$.
For uniqueness, consider a general state $|\psi\rangle=\sum_{\mathcal C}a_{\mathcal C}|\mathcal C\rangle$ for coefficients $a_{\mathcal C}\in\C$ and dimer configurations $\mathcal C$, and note that a ground state must satisfy $P_A|\psi\rangle=0$ for every projection term in \eqref{eqn:Htri2}. Using ergodicity with $4$ and $6$-cycles from Ref.~\cite{hartarsky2024local}, imposing $P_A|\psi\rangle=0$ then determines all coefficients $a_{\mathcal C}$ starting from a single one of these coefficients, implying the ground state must be unique.

More generally, for any simply connected domain in the triangular lattice, one can ensure ground state uniqueness by adding in the 8-cycle terms to the Hamiltonian, using the result of Ref.~\cite{kenyon1996perfect}.

\section{Review of Kasteleyn's method}

In this section, we briefly review the Kasteleyn matrix and some of its properties. For further references, see e.g. Refs.~\cite{kasteleyn1961statistics,kenyon2009lectures,cimasoni2014geometry,gorin2021lectures}.
For a planar graph, one can always define a Kasteleyn orientation \cite{kasteleyn1961statistics,kasteleyn1967graph,fendley2002classical,cimasoni2014geometry}, which can be defined as an assignment of directions to the edges so that the number of edges oriented counterclockwise around each face is odd \footnote{One could alternatively define a Kasteleyn orientation in terms of clockwise orientation}. 
For each directed edge $(v_1\to v_2)$ between two vertices in the graph, we assign a positive orientation if it agrees with the Kasteleyn orientation and a negative orientation if it disagrees. Recall we consider edge weights $w(v_1,v_2)=w(v_2,v_1)$ assigned to the edges of the graph.
The Kasteleyn matrix $K$ is a weighted, signed adjacency matrix which can be defined as:
\begin{align}\label{eqn:K-def}
K(v_1,v_2)&=\begin{cases}
w(v_1,v_2),&(v_1\to v_2)\text{ has positive orientation}\\
-w(v_1,v_2)&(v_1\to v_2)\text{ has negative orientation}\\
0,&(v_1,v_2)\text{ is not an edge}
\end{cases}.
\end{align}
Note that $K$ is skew-symmetric. If the graph is bipartite, then $K$ has a block structure and one can instead consider half-size Kasteleyn matrices where the rows and columns are indexed by one of the two sets of vertices. Since we work on the triangular lattice, we will however have to use the full size matrix defined in \eqref{eqn:K-def}.

For a dimer configuration $\dconfig$, the weight $\cw(\dconfig)=\prod_{e \in \mathcal{C}} w_{e}$ is the product of all edge weights of dimers in $\dconfig$.
By construction of the Kasteleyn orientation, the partition function can be written as \cite{kasteleyn1961statistics}
\begin{align}
Z\equiv\sum_\dconfig \cw(\dconfig)=|\Pf(K)|,
\end{align}
where $\Pf(K)$ denotes the Pfaffian, defined for a $2n\times2n$ skew-symmetric matrix $A$ as
\begin{align}
\Pf(A)&=\frac{1}{2^nn!}\sum_{\sigma\in S_{2n}}\operatorname{sgn}(\sigma)\prod_{i=1}^nA_{\sigma(2i-1),\sigma(2i)}.
\end{align}
The Pfaffian is related to the determinant as
\begin{align}\label{eqn:pfdet}
\Pf(A)^2&=\det A.
\end{align}

Since the Pfaffian of a Kasteleyn matrix counts weighted dimer coverings for the corresponding graph, it can be used to calculate dimer probabilities and correlations. To count the number of dimer configurations where a certain set of edges $E=\{e_1,\ldots,e_k\}$, $e_i=(v_i,u_i)$, appear, one takes the Pfaffian of the Kasteleyn matrix for the subgraph with all vertices $V_E=\{v_1,u_1,v_2,u_2,\ldots,v_k,u_k\}$ (and their attached edges) removed. This leads to the Pfaffian form of Kenyon's formula \cite{kenyon2014conformal}, \cite[\S6.3]{cimasoni2007dimers},
\begin{align}\label{eqn:kenyon}
\P[e_1,\ldots,e_k\text{ all occur in a dimer covering}]&=(-1)^k\left[\prod_{i=1}^kK(v_i,u_i)\right]\Pf(K^{-1}_{V_E}),
\end{align}
where the Pfaffian is over the $2|E|\times 2|E|$ submatrix of $K^{-1}$ corresponding to the $2|E|$ vertices in $V_E$. The product $\prod_{i=1}^kK(v_i,u_i)$ is just a signed product of the weights for the edges in $E$.

As an example, \eqref{eqn:kenyon} gives the dimer-dimer correlator between dimers $\mu=(\mu_0,\mu_1)$ and $\nu=(\nu_0,\nu_1)$ as 
\begin{align}\label{eqn:dimer-dimer}
\langle d_\mu d_\nu\rangle-\langle d_\mu\rangle\langle d_\nu\rangle&=K(\mu_0,\mu_1)K(\nu_0,\nu_1)\left[-K^{-1}(\mu_0,\nu_0)K^{-1}(\mu_1,\nu_1)+K^{-1}(\mu_0,\nu_1)K^{-1}(\mu_1,\nu_0)\right].
\end{align}
Note this formula has an additional term compared to the bipartite case, where the dimer-dimer correlator is proportional to a single product of the Green's function at two points \cite{kenyon2006dimers}.
Equation~\eqref{eqn:kenyon} is also used to derive the vison correlator expression \mainvisonK\ in the Main Text following \cite[\S II.B]{shah2025breakdown} and using \eqref{eqn:pfdet}.

Kasteleyn's method can also be used to count dimer configurations on the torus using four Pfaffians \cite{kasteleyn1961statistics}, which we will use in the next section.

\section{Infinite domain limit}
\begin{figure}[htb]
\includegraphics{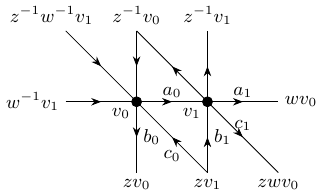}
%\begin{tikzpicture}[scale=1.2,baseline=(current bounding box.center)]
%\def\wcol{black} % color for weighted edge + label
%\def\ocol{black} % color for unweighted edge labels
%\begin{scope}[decoration={
%    markings,
%    mark=at position 0.5 with {\arrow{Stealth}}}
%    ] 
%\foreach \x in {-1,1}{ % horizontal edges, but not the weighted one
%\draw[postaction={decorate}](\x,0)--(\x+1,0);
%}
%\draw[postaction={decorate},color=\wcol] (0,0)--(1,0); % weighted horizontal edge
%
%\foreach \y in {0,1}{
%\draw[postaction={decorate}](0,\y)--(0,\y-1); % vertical edges
%\draw[postaction={decorate}](1,\y-1)--(1,\y); % vertical edges
%\draw[postaction={decorate}](1,\y-1)--(0,\y-0);
%}
%\draw[postaction={decorate}](1,0)--(2,-1);
%\draw[postaction={decorate}](-1,1)--(0,0);
%\end{scope}
%\def\rad{.07}
%\draw[fill=black] (0,0) circle (\rad);
%\draw[fill=black] (1,0) circle (\rad);
%
%\node[below left] at (0,0) {$v_0$};
%\node[below left] at (1,0) {$v_1$};
%
%\node[above,color=\wcol] at (.5,-.03) {$a_0$};
%\begin{scope}[color=\ocol]
%\node[above] at (1.5,-.03) {$a_1$};
%\node[above] at (1.5,-.5) {$c_1$};
%\node[right] at (1,-.5) {$b_1$};
%\node[right] at (0,-.5) {$b_0$};
%\node[below] at (.5,-.6) {$c_0$};
%\end{scope}
%
%\node[left] at (-1,0) {$w^{-1}v_1$};
%\node[right] at (2,0) {$wv_0$};
%\node[above] at (0,1) {$z^{-1}v_0$};
%\node[above] at (1,1) {$z^{-1}v_1$};
%\node[below] at (0,-1) {$zv_0$};
%\node[below] at (1,-1) {$zv_1$};
%\node[above] at (-1.2,1) {$z^{-1}w^{-1}v_1$};
%\node[below] at (2,-1) {$zwv_0$};
%\end{tikzpicture}
\caption{Fundamental domain for the $6$-parameter $2\times1$ periodic triangular lattice. The horizontal weights are $a_0$ and $a_1$, vertical weights are $b_0$ and $b_1$, and diagonal weights are $c_0$ and $c_1$. As in the Main Text, horizontal and vertical translates of the fundamental domain are indexed by variables $w,z\in\C$. }\label{fig:fd6}
\end{figure}

In this section, we analytically derive the double integral formulas for $K^{-1}$ in the infinite domain periodic limit. For the general 6-parameter model on the triangular lattice (Fig.~\ref{fig:fd6}), we use this to show that there is exponential decay of the dimer-dimer correlator off of the set where one of the horizontal or diagonal edge weights is equal to the sum of the other three remaining such weights (Theorem~\ref{thm:dimer-dimer}).
The general method is similar to that of the bipartite case \cite{cohn2001variational,kenyon2006dimers,gorin2021lectures}, using block diagonalization of the Kasteleyn matrix.
In this section, to avoid confusion with a complex parameter $w\in\C$, we will use $W(e)$ to represent the weight of an edge $e$.

We start in the general setting of a graph, which may be non-bipartite, on the torus.
Considering periodic boundary conditions, or a graph $G$ on the torus, the partition function is given by a linear combination of four Pfaffians \cite{kasteleyn1961statistics}. There are four Kasteleyn matrices $K^{\alpha\beta}$ with $\alpha,\beta\in\{0,1\}$, defined by constructing the usual Kasteleyn matrix and then flipping the sign on horizontal ($\alpha=1$) and/or vertical ($\beta=1$) edges that wrap around the boundaries. There are then $s_{\alpha\beta}\in\{-1,1\}$ such that the partition function is given by
\begin{align}\label{eqn:Z}
Z(G)&=\frac{1}{2}\left[s_{00}\Pf(K^{00})+s_{01}\Pf(K^{01})+s_{10}\Pf(K^{10})+s_{11}\Pf(K^{11})\right].
\end{align}
It was shown in \cite{cimasoni2007dimers} that the following formula for the correlations holds: Letting $V_E=\{u_1,v_1,\ldots,u_k,v_k\}$ be a set of verticies in $G$, then
\begin{align}\label{eqn:correlations}
\E[\oneb_{u_1v_1}\cdots\oneb_{u_kv_k}]&=\frac{(-1)^k}{Z(G)}\sum_{\alpha\beta}s_{\alpha\beta}\Pf(K^{\alpha\beta})\Pf((K^{\alpha\beta})^{-1}_{V_E})\prod_{i=1}^k \epsilon_{\alpha\beta}(u_i,v_i)w(u_i,v_i),
\end{align}
where $A_E$ denotes the restriction of the matrix $A$ to rows and columns with indices in $E$, and $\epsilon_{\alpha\beta}(u_\ell,v_\ell)$ denotes the $\alpha\beta$-Kasteleyn orientation of the directed edge $(u_\ell,v_\ell)$; i.e. $\epsilon_{\alpha\beta}(u_\ell,v_\ell)=1$ if edge $(u_\ell,v_\ell)$ aligns with the $\alpha\beta$-Kasteleyn orientation, $\epsilon_{\alpha\beta}(u_\ell,v_\ell)=-1$ if the edge is in the opposite orientation, and $0$ otherwise.

Now suppose $\Gamma$ is a $\Z^2$-periodic planar graph, and let $\Gamma_{mn}=\Gamma/(m\Z\times n\Z)$, which we will view as a graph embedded in the torus. We will assume that the edges of $\Gamma$ have a positive doubly periodic weight function $w$, and that $(m,n)$ are chosen compatible with the periods.
Select a fundamental domain $\Gamma^{0,0}_{mn}$ for $\Gamma_{mn}$ (for example, see Fig.~\ref{fig:fd6} for a fundamental domain for the $2\times1$ periodic triangular lattice) and let $\Gamma_{mn}^{i,j}$ be the translate of $\Gamma_{mn}^{0,0}$ by $(i,j)$. Similarly, for a point $u\in\Gamma_{mn}^{0,0}$ in the fundamental domain, we will let $u^{i,j}$ denote its translate by $(i,j)$.

For $\alpha,\beta\in\{0,1\}$, let $W^{\alpha\beta}$ be the block 2D Fourier matrix indexed by vertices of $\Gamma_{mn}$ defined as
\begin{align}
\langle u^{i,j}|W^{\alpha\beta}|v^{k,\ell}\rangle&=\begin{cases}\frac{1}{\sqrt{mn}}\omega^{-(2i+\alpha)k}\zeta^{-(2j+\beta)\ell},&u=v\\
0,&\text{otherwise}
\end{cases},
\end{align}
where $\omega=e^{\pi i/m},\zeta=e^{\pi i/n}$ are the $(2m)$th and $(2n)$th roots of unity respectively. Let $K^{\alpha\beta}_{mn}$, $\alpha,\beta\in\{0,1\}$, be the four Kasteleyn matrices for $\Gamma_{mn}$ on the torus.
\begin{lem}[block diagonalization]\label{lem:block}
The matrix
\begin{align}
L&=W^{\alpha\beta}K_{mn}^{\alpha\beta}(W^{\alpha\beta})^{-1}
\end{align}
is block diagonal with respect to the partitioning of $\Gamma_{mn}$ into subsets $\Gamma_{mn}^{i,j}$. Furthermore, the diagonal block correspnding to $\Gamma_{mn}^{i,j}$ is given by $L_{\omega^{2i+\alpha},\zeta^{2j+\beta}}$ where
\begin{align}\label{eqn:Lwz-gen}
\langle u|L_{w,z}|v\rangle&=\sum_{i'=-1}^1\sum_{j'=-1}^1w^{i'}z^{j'}\langle u|K_{mn}^{00}|v^{i',j'}\rangle.
\end{align}
\end{lem}
\begin{proof}
Inverting the Fourier matrix gives
\begin{align}
(W^{\alpha,\beta})^{-1}(u^{i,j},v^{k,\ell})&=\begin{cases}\frac{1}{\sqrt{mn}}\omega^{(2k+\alpha)i}\zeta^{(2\ell+\beta)j},&u=v\\
0,&\text{otherwise}\end{cases}.
\end{align}
Let $u,v\in\Gamma^{0,0}_{mn}$ be points in the fundamental domain. Then for all $i,k\in\{0,\ldots,m-1\}$ and $j,\ell\in\{0,\ldots,n-1\}$, one has
\begin{align}\label{eqn:blockmultiplication}
\langle u^{i,j}|W^{\alpha\beta}K_{mn}^{\alpha\beta}(W^{\alpha\beta})^{-1}|v^{k,\ell}\rangle&=\frac{1}{mn}\sum_{i',k'=0}^{m-1}\sum_{j',\ell'=0}^{n-1}\omega^{-(2i+\alpha)i'+(2k+\alpha)k'}\zeta^{-(2j+\beta)j'+(2\ell+\beta)\ell'}\langle u^{i',j'}|K_{mn}^{\alpha\beta}|v^{k',\ell'}\rangle.
\end{align}
Consider all terms in the sum corresponding to a $\Z^2$ translate of a pair $(u^{i',j'},v^{k',\ell'})$; i.e. where $(i',j')$ and $(k',\ell')$ are both shifted by some $(r,s)\in\Z^2$ (and taken modulo $m\Z\times n\Z$). The value of
\begin{align}
\omega^{\alpha(k'-i')}\zeta^{\beta(\ell'-j')}\langle u^{i',j'}|K_{mn}^{\alpha\beta}|v^{k',\ell'}\rangle
\end{align}
will be constant across all such translates; the only place of concern is if the translate crosses the border of the domain, but then $\omega^{\alpha(k'-i')}\zeta^{\beta(\ell'-j')}$ will undergo the same sign change as the element of the Kasteleyn matrix. The remaining portion of the terms in \eqref{eqn:blockmultiplication} which changes across these translates is then 
\[
\omega^{2(kk'-ii')}\zeta^{2(\ell\ell'-jj')}.
\]
Since the sum over the translates of this expresion is $0$ unless $i=k$ and $j=\ell$, it follows that $W^{\alpha\beta}K_{mn}^{\alpha\beta}(W^{\alpha\beta})^{-1}$ is indeed block diagonal. 

To obtain \eqref{eqn:Lwz-gen}, note that when $i=k$ and $j=\ell$, each term in the sum in \eqref{eqn:blockmultiplication} is invariant with respect to $\Z^2$ translations of $(i',j')$ and $(k',\ell')$ together, and thus
\begin{align}
\langle u^{i,j}|W^{\alpha\beta}K_{mn}^{\alpha\beta}(W^{\alpha\beta})^{-1}|v^{i,j}\rangle&=\sum_{i'=0}^{m-1}\sum_{j'=0}^{n-1}\omega^{(2i+\alpha)i'}\zeta^{(2j+\beta)j'}\langle u|K_{mn}^{\alpha\beta}|v^{i',j'}\rangle.
\end{align}
Since $u\in\Gamma^{0,0}_{mn}$, the Kasteleyn matrix entry can only be nonzero for $i'\in\{0,1,m-1\}$ and $j'\in\{0,1,n-1\}$. As the signs in the $i'=m-1$ or $j'=n-1$ cases cancel with those arising from the orientation $\epsilon_{\alpha\beta}$, this is precisely the desired expression.
\end{proof}

\begin{cor}[finite domain expressions]\label{cor:finite}
Let $L_{w,z}$ be defined as in \eqref{eqn:Lwz-gen} and set $P(w,z)=\det L_{w,z}$, which is the characteristic polynomial. Then
\begin{align}
\det K_{mn}^{\alpha\beta}&=\prod_{i=1}^{m-1}\prod_{j=1}^{n-1}P(\omega^{2i+\alpha},\zeta^{2j+\beta}),
\end{align}
where $\omega,\zeta$ are $(2m)$th, $(2n)$th roots of unity respectively. Furthermore,
\begin{align}\label{eqn:Kfinite}
\langle u^{i,j}|(K_{mn}^{\alpha\beta})^{-1}|v^{k,\ell}\rangle&=\frac{1}{mn}\sum_{s=0}^{m-1}\sum_{t=0}^{n-1}\omega^{-(2s+\alpha)(k-i)}\zeta^{-(2t+\beta)(\ell-j)}\langle u|L^{-1}_{\omega^{2s+\alpha},\zeta^{2t+\beta}}|v\rangle.
\end{align}
\end{cor}

\begin{thm}[infinite domain limit]\label{thm:inflimit}
Let $L_{w,z}$ be defined as in Corollary~\ref{cor:finite}, and let $\T^1$ be the unit circle in $\C$. 
Then as $m,n\to\infty$, each $(K_{mn}^{\alpha\beta})^{-1}$ converges elementwise to $K^{-1}$ given by
\begin{align}\label{eqn:Kinfinite}
\langle u^{i,j}|K^{-1}|v^{k,\ell}\rangle&=-\frac{1}{4\pi^2}\int_{\T^1}\int_{\T^1}\langle u|L_{w,z}^{-1}|v\rangle\,w^{-(k-i)}z^{-(\ell-j)}\frac{dw}{w}\frac{dz}{z}.
\end{align}
Additionally, for edges $e_1=(u_1,v_1),\ldots,e_k=(u_k,v_k)$, letting $V_E=\{u_1,v_1,\ldots,u_k,v_k\}$,
\begin{align}\label{eqn:corr_infinite}
\lim_{m,n\to\infty}\E[\oneb_{e_1}\cdots\oneb_{e_k}]&=(-1)^k\left(\prod_{i=1}^k\epsilon_{00}(u_i,v_i)W(u_i,v_i)\right)\Pf(K^{-1})_{V_E}.
\end{align}
\end{thm}
\begin{proof}
For \eqref{eqn:Kinfinite}, note that the right side of \eqref{eqn:Kfinite} is a Riemann sum converging to the expression in \eqref{eqn:Kinfinite} for $\langle u^{i,j}|K^{-1}|v^{k,\ell}\rangle$.
For \eqref{eqn:corr_infinite}, note that for sufficiently large domains $\Gamma_{mn}$, we may use \eqref{eqn:correlations} with $\epsilon_{\alpha\beta}$ replaced by $\epsilon_{00}$ since none of the selected edges will cross the boundaries of the domain.  Then we have for sufficiently large $m,n$,
\begin{align}
\E[\oneb_{e_1}\cdots\oneb_{e_k}]&=(-1)^k\sum_{\alpha\beta}\frac{s_{\alpha\beta}\Pf(K_{mn}^{\alpha\beta})}{Z_{mn}}\Pf((K_{mn}^{\alpha\beta})^{-1}_E)\prod_{i=1}^k \epsilon_{00}(u_i,v_i)W(u_i,v_i).
\end{align}
The coefficients $\frac{s_{\alpha\beta}\Pf(K_{mn}^{\alpha\beta})}{Z_{mn}}$ are bounded and sum to one by \eqref{eqn:Z}. Then using \eqref{eqn:Kinfinite} gives \eqref{eqn:corr_infinite}.
\end{proof}

We now specialize to the $2\times1$ periodic triangular lattice, whose fundamental domain is drawn in Fig.~\ref{fig:fd6}.
\begin{thm}[dimer-dimer correlator for the $2\times1$ triangular lattice]\label{thm:dimer-dimer}
Consider the $2\times1$ periodic triangular lattice $\Gamma$ whose fundamental domain is drawn in Fig.~\ref{fig:fd6}.
Let $e_1,e_2$ be two edges in $\Gamma$, and let $d_{e}$ be the dimer occupation number of the edge $e$. Define $e^{r,s}$ the edge $(r,s)+e_2$.
If none of the weights $a_0,a_1,c_0,c_1$ are a sum of the other three, then the dimer-dimer correlator between edges $e_1$ and $e_2^{r,s}$ decays exponentially as $|r|,|s|\to\infty$.
\end{thm}
\begin{proof}
From the fundamental domain drawn in Fig.~\ref{fig:fd6}, we can read off the matrix $L_{w,z}$ as
\begin{align}\label{eqn:Lwz}
L_{w,z}&=\begin{bmatrix}
b_0(z-z^{-1})&a_0-a_1w^{-1}-c_0z-c_1(wz)^{-1}\\
-a_0+a_1w+c_0z^{-1}+c_1wz& -b_1(z-z^{-1})
\end{bmatrix},
\end{align}
and thus
\begin{align}
P(w,z)\equiv\det L_{w,z}&=-b_0b_1(z-z^{-1})^2+(a_0-a_1w^{-1}-c_0z-c_1(wz)^{-1})(a_0-a_1w-c_0z^{-1}-c_1wz).
\end{align}
For $w,z\in\T^1$, both terms in the sum above are equal to the magnitude squared of a complex number.
Thus $P(w,z)$ can only vanish if both complex numbers vanish, which corresponds precisely to the case when one of $a_0,a_1,c_0,c_1$ is a sum of the other three. Consequently, if this condition does not hold, then for any choice of $u,v$, $L_{w,z}^{-1}(u,v)$ is a smooth and bounded function over $w,z\in\T^1$.

Now, let $e_1=(u_1,v_1)$ and $e_2=(u_2,v_2)$. By Eq.~\eqref{eqn:corr_infinite} of Theorem~\ref{thm:inflimit}, the dimer-dimer correlator is given by
\begin{align}
\operatorname{Cov}(d_{e_1},d_{e_2^{r,s}})&=\left(\prod_{i=0}^1\epsilon_{00}(u_i,v_i)W(u_i,v_i)\right)\left[\Pf(K^{-1})_{V_E}-\Pf(K^{-1})_{V_{E_1}}\Pf(K^{-1})_{V_{E_2}}\right],
\end{align}
where $V_{E_1}=\{u_1,v_1\}$, $V_{E_2}=\{u_2^{r,s},v_2^{r,s}\}$, and $V_E=V_{E_1}\cup V_{E_2}$. Expanding out the Pfaffians one gets
\begin{align*}
\Pf(K^{-1})_{V_E}-\Pf(K^{-1})_{V_{E_1}}\Pf(K^{-1})_{V_{E_2}}&=-K^{-1}(u_1,u_2^{r,s})K^{-1}(v_1,v_2^{r,s})+K^{-1}(u_1,v_2^{r,s})K^{-1}(v_1,u_2^{r,s}).
\end{align*}
The function $L^{-1}_{w,z}(u,v)$ is a ratio of trigonometric polynomials and is analytic when $P(w,z)$ has no zeros. Since the formula \eqref{eqn:Kinfinite} for $K^{-1}$ is a Fourier transform of $\langle u|L_{w,z}^{-1}|v\rangle$, a Paley--Wiener theorem implies these terms and the above correlator decay exponentially in $r,s$.
\end{proof}

\section{Exponential decay and correlation lengths off the critical point}\label{sec:expdecay}

In this section, we consider the $2\times1$ periodic triangular lattice with one horizontal edge weight $\wt$, and the other five edge weights equal to one. We first explicitly write the formulas from the previous section in this case, and then use contour integration and a Paley--Wiener theorem to analytically study the correlation length of $K^{-1}$ along a vertical path as $\wt\to3$. We also compare to numerical plots of the correlation lengths.

To obtain the explicit formulas for the double integrals, we start from \eqref{eqn:Lwz} which gives the magnetic Kasteleyn matrix
\begin{align}\label{eqn:L}
L_{w,z}&=\begin{bmatrix}
z-z^{-1} & \wt-z-w^{-1}-z^{-1}w^{-1}\\
-\wt+w+z^{-1}+zw & -(z-z^{-1})\\
\end{bmatrix}.
\end{align}
The inverse is
\begin{align}\label{eqn:Kinv}
L_{w,z}^{-1}&=\frac{1}{P(w,z)}\begin{bmatrix}
-(z-z^{-1})&-\wt+z+w^{-1}+z^{-1}w^{-1}\\
\wt-w-z^{-1}-zw &z-z^{-1}\\
\end{bmatrix},
\end{align}
where $P(w,z)=\det L_{w,z}=-(z-z^{-1})^2+(\wt-z-w^{-1}-z^{-1}w^{-1})(\wt-w-z^{-1}-zw)$ is the \emph{characteristic polynomial}. When $w=e^{ik_x}$ and $z=e^{ik_y}$ are on the unit circle, this gives
\begin{align}\label{eqn:Pwz}
P(e^{ik_x},e^{ik_y})&=4\sin^2k_y+|\wt-e^{ik_y}-e^{-ik_x}-e^{-ik_x}e^{-ik_y}|^2,
\end{align}
which is also the product of  band dispersions for the corresponding free-fermion model.

We now index vertices in the triangular lattice by coordinates $(x,y,t)$, where $(x,y)\in\Z^2$ denotes the translation of the fundamental cell, and $u,v\in\{0,1\}$ specifies the vertex $v_0$ or $v_1$ in Fig.~\ref{fig:fd6} or in Fig.~\maintriangle\ of the Main Text.
Then in the infinite size periodic limit, \eqref{eqn:Kinv} and Theorem~\ref{thm:inflimit} imply
\begin{align}\label{eqn:Kinvinf}
\langle 0,0,u|K^{-1}|x,y,v\rangle&= \int_0^{2\pi}\int_0^{2\pi}\frac{Q_{uv}(e^{ik_x},e^{ik_y})e^{-i(k_xx+k_yy)}}{4\sin^2k_y+|\wt-e^{ik_y}-e^{-ik_x}-e^{-ik_x}e^{-ik_y}|^2} \frac{dk_x}{2\pi}\frac{dk_y}{2\pi},
\end{align}
where $Q_{uv}(w,z)$ is the rational function given by the $(u,v)$ matrix entries [without $1/P(w,z)$] of \eqref{eqn:Kinv}.

To study exponential decay and correlation lengths analytically, we consider the infinite size periodic limit as in the previous section. 
The matrix elements of $K^{-1}$ given in \cref{eqn:Kinvinf} correspond to the 2D Fourier transform of a rational function $F$ whose denominator is given by the product of band dispersions $P(e^{ik_x},e^{ik_y})$ in \cref{eqn:Pwz}. 
As in Theorem~\ref{thm:dimer-dimer}, using that $P(e^{ik_x},e^{ik_y})$ is a sum of two nonnegative terms, we see it can only be zero if $k_y=k_x=0$ and $\wt=3$.
In particular, for $\wt\ne3$, there is a band gap since $P(e^{ik_x},e^{ik_y})\ne0$ for all $k_x,k_y\in[0,2\pi]$, and as follows in Theorem~\ref{thm:dimer-dimer}, analyticity of the resulting function $F$ implies exponential decay of its Fourier transform, and hence of the matrix elements of $K^{-1}$.
As a result, we know from the above argument expressed in Theorem~\ref{thm:dimer-dimer} that the dimer-dimer connected correlator decays exponentially for $\wt\ne3$,
\begin{align}
|\langle d_\mu d_\nu\rangle-\langle d_\mu\rangle\langle d_\nu\rangle|&\le Ce^{-c\operatorname{dist}(\mu,\nu)},
\end{align}
where $C,c$ may depend on $\wt$, and $\operatorname{dist}(\mu,\nu)$ denotes some distance (such as Euclidean distance) on the lattice.

Next, for $\wt\ne3$, we calculate the correlation length $\xi_\mathrm{G}$ for the inverse Kasteleyn limit, which corresponds to the real-space Green's function in the corresponding free-fermion model, along a vertical path.
We start with the infinite size periodic limit formula \eqref{eqn:Kinvinf}, which gives expressions of the form
\begin{multline}\label{eqn:dcontour}
\int_0^{2\pi}\int_0^{2\pi}\frac{e^{ij k_x}e^{i\ell k_y}}{4\sin^2k_y+|\wt-e^{ik_y}-e^{-ik_x}-e^{-ik_x}e^{-ik_y}|^2}\frac{dk_x}{2\pi}\frac{dk_y}{2\pi}\\
=\frac{1}{(2\pi i)^2}\oint_{S^1}\oint_{S^1}\frac{w^jz^\ell}{c_1(z,\wt)w^2+c_2(z,\wt)w+c_3(z,\wt)}\,dw\,\frac{dz}{z},
\end{multline}
where
\begin{align}
\begin{aligned}
c_1(z,\wt)&=(z-\wt)(z+1)\\
c_2(z,\wt)&=5+\wt^2-(z^2+z^{-2})+(1-\wt)(z+z^{-1})\\
c_3(z,\wt)&=(z^{-1}-\wt)(z^{-1}+1)
\end{aligned}
\end{align}
The denominator of the integrand factors as $c_1(z,\wt)w^2+c_2(z,\wt)w+c_3(z,\wt)=c_1(z,\wt)[w-r_+(z,\wt)][w-r_-(z,\wt)]$ for
\begin{align}
r_\pm(z,\wt)&= \frac{-c_2(z,\wt)\pm\sqrt{c_2(z,\wt)^2-4c_1(z,\wt)c_3(z,\wt)}}{2c_1(z,\wt)},
\end{align}
for $z\ne-1,\wt$ \footnote{If $z\in\{-1,\wt\}$, then $c_1(z,\wt)=0$ and the denominator is simply linear with a zero at $w=-c_3(z,\wt)/c_2(z,\wt)=0$. The contour integral over $w$ is then zero for $\ell,j\ne0$, which agrees with the formulas \eqref{eqn:single-int} and \eqref{eqn:fa} for $z\ne-1,\wt$.}.
For $z=e^{ik_y}$ with $k_y\in\R$, we have
\begin{align}
|r_+(z,\wt)r_-(z,\wt)|&=\frac{\left|c_2(z,\wt)^2-\left(c_2(z,\wt)^2-4c_1(z,\wt)c_3(z,\wt)\right)\right|}{4|c_1(z,\wt)|^2}=\frac{|c_3(z,\wt)|}{|c_1(z,\wt)|}=1.
\end{align}
Also, $c_2(e^{ik_y},\wt)$ is real and is $>0$, so we see that $|r_+(e^{ik_y},\wt)|\le|r_-(e^{ik_y},\wt)|$. We know there are no zeros of the denominator of the integrand in \eqref{eqn:dcontour} for any $|z|=|w|=1$ iff $\wt\ne 3$. Thus for $\wt\ne3$, we must have $|r_+(z,\wt)|<1$ and $|r_-(z,\wt)|>1$ for all $|z|=|w|=1$. Therefore, for $j\ge0$ and $\wt\ne3$,  the only pole inside the unit circle is at $w=r_+(z,\wt)$, and the residue theorem implies
\begin{align}
\nonumber\frac{1}{(2\pi i)^2}\oint_{S^1}\frac{z^\ell\,dz}{z}\oint_{S^1}\frac{w^j}{c_1(z,\wt)[w-r_+(z,\wt)][w-r_-(z,\wt)]}\,dw&=\int_0^{2\pi}\frac{\left[r_+(e^{ik_y},\wt)\right]^j\,e^{i\ell k_y}}{c_1(z,\wt)[r_+(e^{ik_y},\wt)-r_-(e^{ik_y},\wt)]}\frac{dk_y}{2\pi}\\
&=\int_0^{2\pi}\frac{\left[r_+(e^{ik_y},\wt)\right]^j\,e^{i\ell k_y}}{\sqrt{c_2(e^{ik_y},\wt)^2-4c_1(e^{ik_y},\wt)c_3(e^{ik_y},\wt)}}\frac{dk_y}{2\pi}.\label{eqn:single-int}
\end{align}

We consider the decay along a vertical path in the triangular lattice. 
For example, we can consider fixed $j\ge0$, and replace $e^{i\ell k_y}$ in the numerator with $\sin(k_y)e^{i\ell k_y}$; up to constant factors this corresponds to the Green's function between two points separated vertically.
The integral \eqref{eqn:single-int}, with modified numerator, is then the Fourier transform of the function
\begin{align}\label{eqn:fa}
f_a(k_y)&=\frac{\sin(k_y)\left[r_+(e^{ik_y},\wt)\right]^j}{\sqrt{c_2(e^{ik_y},\wt)^2-4c_1(e^{ik_y},\wt)c_3(e^{ik_y},\wt)}},
\end{align}
for $k_y\in \R/(2\pi\Z)$. 
The rate of decay of the Fourier transform of $f_a$ is determined by the nearest complex zero of $f_a$, due to a Paley--Wiener theorem: 
\begin{lem}\label{lem:fourier}
If $f$ is $2\pi$-periodic and real analytic with analytic extension to a neighborhood of the strip $\{z\in\C:|\im z|\le\kappa\}$ for some $\kappa>0$, then the Fourier coefficients $\hat{f}(\xi)=\frac{1}{2\pi}\int_0^{2\pi}f(x)e^{-i\xi x}\,dx$ satisfy
\begin{align}
|\hat{f}(\xi)|\le C e^{-\kappa|\xi|},
\end{align}
for some constant $C=C(f)$. Conversely, if the Fourier coefficients have the above exponential decay, then $f$ has analytic extension to  the strip $\{z\in\C:|\im z|<\kappa\}$.
\end{lem}
More generally, a $d$-dimensional version can be formulated, see for example, Ref.~\cite[Lemma 5.6]{broer2011dynamical}, \cite{arnold1987geometrical}.

Therefore to determine the inverse correlation length of $K^{-1}$, we locate the zeros of
\[
d(e^{ik_y},\wt):=\sqrt{c_2(e^{ik_y},\wt)^2-4c_1(e^{ik_y},\wt)c_3(e^{ik_y},\wt)},
\]
for $k_y=t+is$ which are closest to the real line.

\begin{figure}[htb]
\includegraphics[width=.4\textwidth]{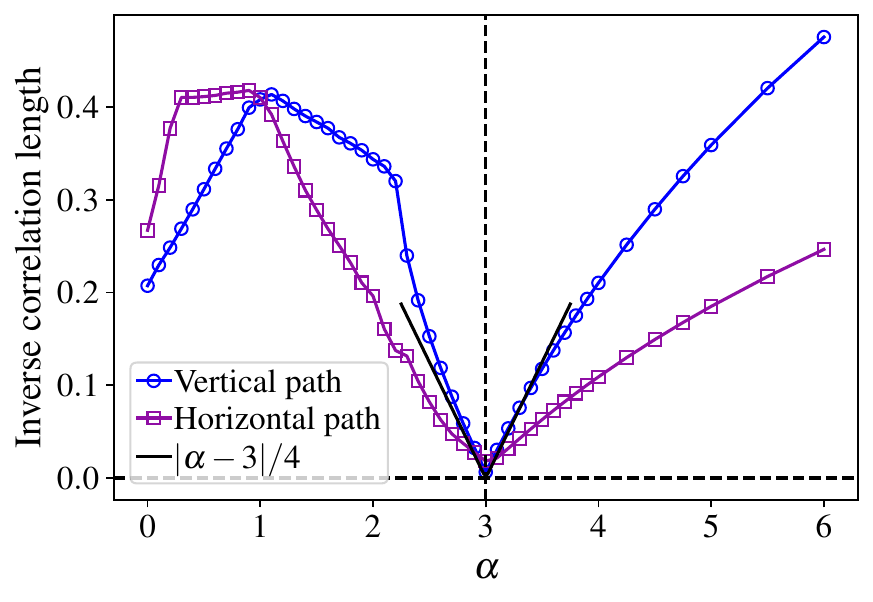}
\caption{Inverse correlation lengths $\xi_G^{-1}$ for the inverse Kasteleyn matrix values $|K^{-1}(v_0,v_\ell)|$ (or real-space Greens' function) along a vertical and horizontal path, using lattice size $303\times303$. 
The correlation length is calculated with respect to vertices $v_\ell$ which include all of those in the zigzag path drawn in Fig.~\mainpath, in particular alternating between columns (for the vertical path) or rows (for the horizontal path). There are two vertices per lattice square in the zigzag path, accounting for the extra factor of $1/2$ in the comparison to $|\wt-3|/4$.
The inverse correlation lengths are calculated using a best fit regression on vertices $\ell=40$ to $150$, as measured from the center $\ell=0$ and ending roughly halfway to the edge at the endpoint $\ell=150$.
}\label{fig:kinvcorr}
\end{figure}

When $\wt=3$, we see that $k_y=0$ solves $d(e^{ik_y},\wt)^2=0$. When we perturb $\wt$ away from 3, this root, which is quadratic for $d(e^{ik_y},\wt)^2$, will split and move into the complex plane. 
We look for these roots by Taylor expanding $d(e^{i(t+is)},\wt)^2$ as $s,t\to0$ and $\wt\to3$, which gives to leading orders,
\begin{align}
d(e^{i(t+is)},\wt)^2&\asymp(64t^2+128its-64s^2-160t^2s^2)[1+(\wt-3)]+(16+24t^2+48its-24s^2+18t^2s^2)(\wt-3)^2.
\end{align}
Taking $t=0$ so the imaginary part of the right hand side is zero,
we solve for the smallest zeros of
\begin{align}
d(e^{-s},\wt)^2&=-64s^2-64s^2(\wt-3)+(16-24s^2)(\wt-3)^2+O(s^3)+O((\wt-3)^3).
\end{align}
Dropping higher order terms gives that this is equal to zero for
\begin{align}
s^2&=\frac{2(\wt-3)^3}{8+8(\wt-3)+3(\wt-3)^2}=\frac{(\wt-3)^2}{4}(1+O(\wt-3)).
\end{align}
Moreover, the numerator of \eqref{eqn:fa} is bounded away from zero, since $\sin(k_y)$ has no $\wt$ dependence, and the other numerator terms were nonzero for $t=s=0$ and $\wt=3$.
Thus the distance to the nearest complex zero of $d(e^{ik_y},\wt)$ for $\wt$ near 3 is $|s|\sim
\frac{1}{2}|\wt-3|$.
This gives the correlation length for $K^{-1}$ (the real-space Green's function in the free-fermion picture) along a vertical path as
\begin{align}\label{eqn:corrKinv}
\xi_\mathrm{G}&\sim\frac{2}{|\wt-3|},\quad \text{ as }\,\wt\to3,
\end{align}
measured in lattice distance. 
One could also study the correlation length along other paths similarly. 
We plot the inverse correlation lengths along different paths numerically in Fig.~\ref{fig:kinvcorr}. In particular, this confirms \eqref{eqn:corrKinv} numerically for the vertical path.

\begin{figure}[htb]
\includegraphics[width=.4\textwidth]{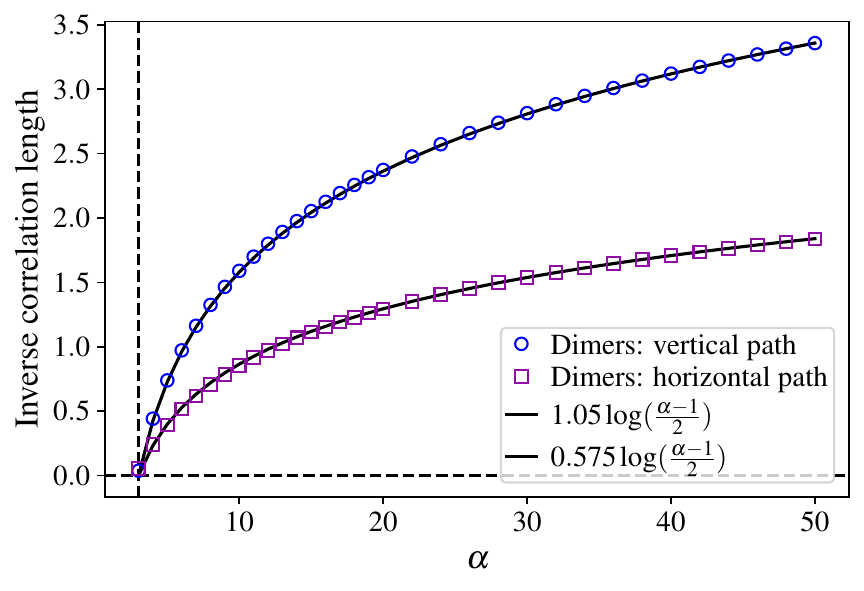}
\caption{Dimer-dimer inverse correlation length from $\wt=3$ to $50$, using lattice size $303\times303$.}\label{fig:largea}
\end{figure}

For large $\wt$, we can show along a horizontal or vertical path that
\begin{align}
\xi_\mathrm{G}\sim \frac{1}{\log(\frac{\wt-1}{2})},\quad\text{ as }\,\wt\to\infty.
\end{align}
As shown in Fig.~\ref{fig:largea}, the dimer-dimer inverse correlation length behaves similarly as $\xi_\mathrm{dimer}\propto\frac{1}{\log(\frac{\wt-1}{2})}$.
Starting from the band dispersion $P(e^{ik_x},e^{ik_y})$ in \eqref{eqn:Pwz}, we see there is a complex zero at $(k_x,k_y)=(0,i\log(\frac{\wt-1}{2}))$. This is also not a zero of any of the $Q_{st}(e^{ik_x},e^{ik_y})$ in \eqref{eqn:Kinv}.
A quick standard estimate shows that if $f:(\R/2\pi\Z)^d\to\C$ has Fourier coefficients $|\hat f(\xi)|\le Ce^{-\kappa\|\xi\|_2}$, then $f$ has analytic extension to the strip $\{x+iv\in(\C/2\pi\Z)^d:\|v\|_2<\kappa\}$. 
Thus for $P(e^{ik_x},e^{ik_y})$, the complex zero prevents analytic extension to $\kappa=\log(\frac{\wt-1}{2})$, which implies $\xi_\mathrm{G}^{-1}\le \log(\frac{\wt-1}{2})$.
To obtain a lower bound on $\xi_\mathrm{G}^{-1}$, let $k_x=\theta+iu$ and $k_y=\varphi+iv$. If $|u|+|v|\le\kappa=(1-\delta)\log(\frac{\wt-1}{2})$, then
\begin{align*}
|P(&e^{i(\theta+iu)},e^{i(\varphi+iv)})|\\
&=\big|e^{2i\theta}e^{-2u}+e^{-2i\theta}e^{2u}-2+(a-e^{-i\varphi}e^v-e^{i\theta}e^{-u}-e^{-i\theta}e^{-i\varphi}e^ue^v)(a-e^{i\varphi}e^{-v}-e^{-i\theta}e^{u}-e^{i\theta}e^{i\varphi}e^{-u}e^{-v})\big|\\
&\ge (\wt-3e^\kappa)^2-2-2e^{2\kappa}=\wt^2-O(\wt^{2-2\delta}).\numberthis
\end{align*}
As $\wt\to\infty$, this is bounded away from zero. Thus there are no complex zeros of $P$ if $|u|+|v|\le (1-\delta)\log(\frac{\wt-1}{2})$. Then along a horizontal or vertical path, a $2$-dimensional version of Lemma~\ref{lem:fourier}, which replaces the absolute values with the $\ell^1$ norm \cite[Lemma 5.6]{broer2011dynamical}, implies $\xi_\mathrm{G}^{-1}\ge(1-\delta)\log(\frac{\wt-1}{2})$ for any $\delta>0$ and sufficiently large $\wt$.

For completeness, we comment on the inverse correlation lengths calculated in Fig.~\maincorr\ in the Main Text. There, the inverse correlation lengths are calculated using linear regression for edge/face distances starting at 30 and ending at 150 (for dimers) or when numerical precision drops below $10^{-15}$ (for vison correlator near $\wt=1$).

\begin{figure}[tb]
\includegraphics[height=1.45in]{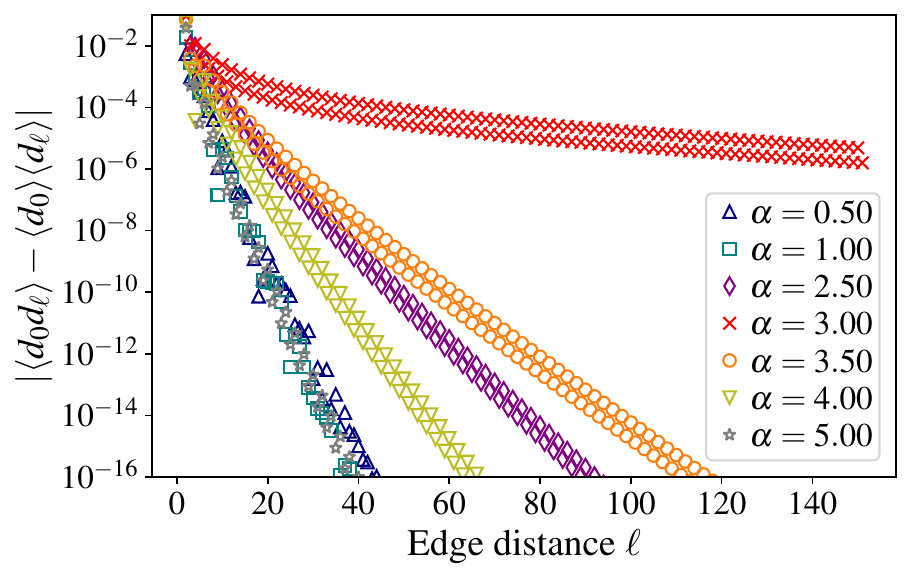}
\includegraphics[height=1.45in]{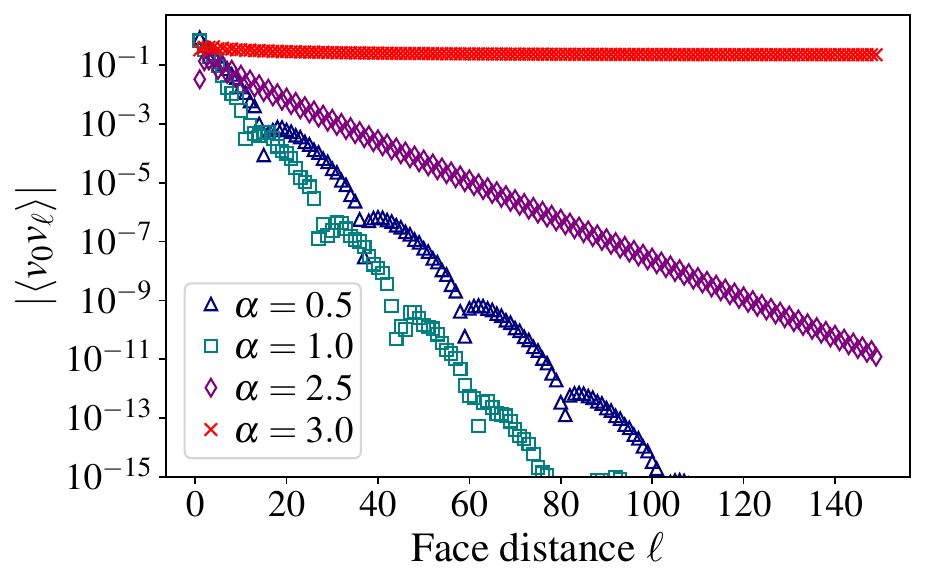}
\includegraphics[height=1.45in]{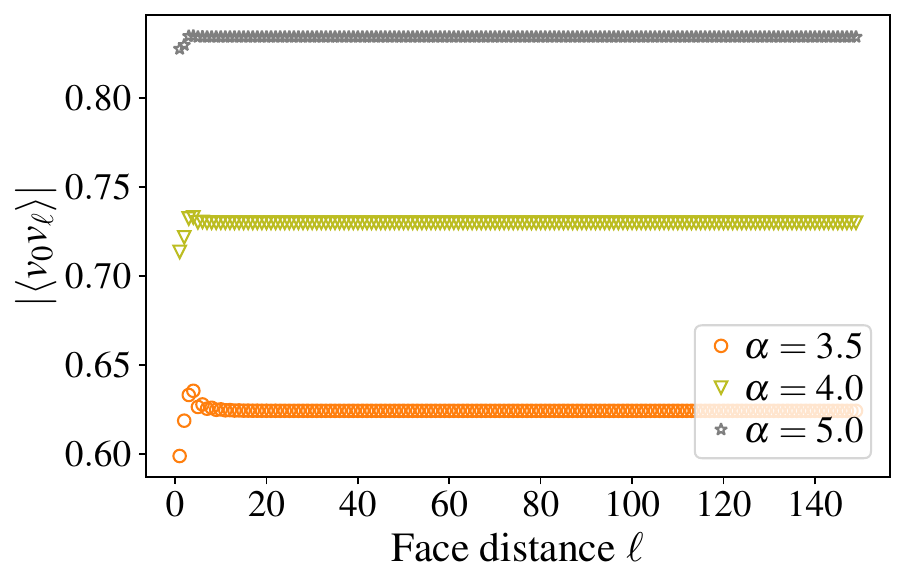}
\caption{
Dimer-dimer and vison correlators for different values of $\wt$, calculated along a vertical path through weight $\wt$ edges on a $303\times303$ size grid. Compare with Fig.~\mainfigcorr\ from the Main Text, which considers a vertical path through weight $1$ edges. The present vertical path is shifted one column over that of Fig.~\mainfigcorr\ of the Main Text.
}\label{fig:visons-vshift}
\end{figure}

\begin{figure}[tb]
\includegraphics[height=1.45in]{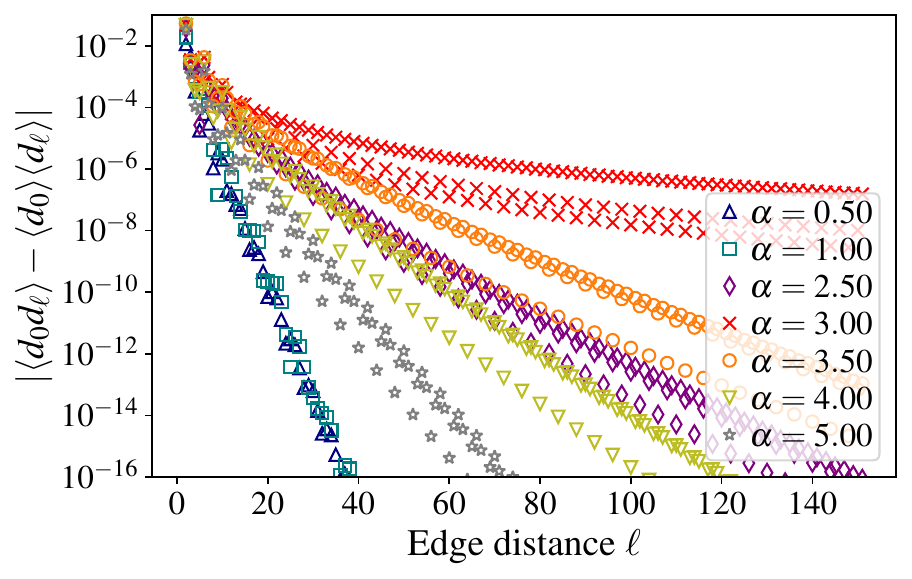}
\includegraphics[height=1.45in]{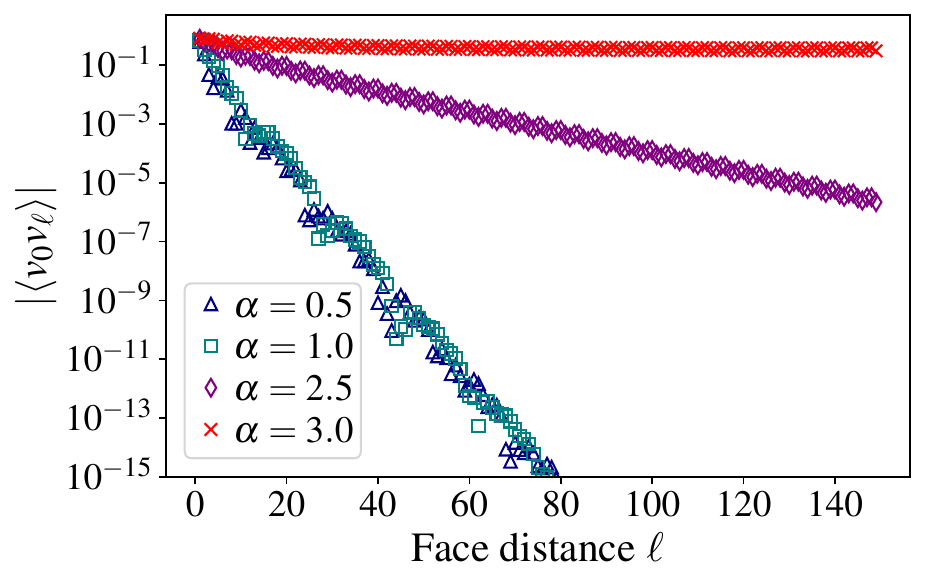}
\includegraphics[height=1.45in]{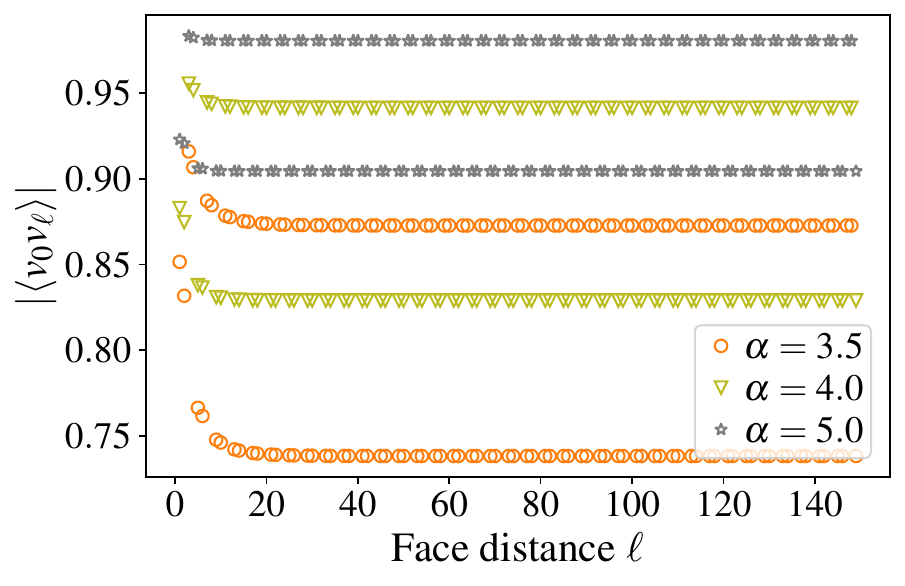}
\caption{
Dimer-dimer and vison correlators for different values of $\wt$, calculated along a horizontal path on a $303\times303$ size grid. Compare with Fig.~\mainfigcorr\ from the Main Text, which considers a vertical path. The correlators shown here for the horizontal path display alternating values depending on whether the path is in a column with horizontal edge weights $\wt$ or $1$.
}\label{fig:visons-hor}
\end{figure}

Finally, we provide additional numerical plots for the dimer-dimer and vison correlators along different paths in Figures~\ref{fig:visons-vshift} and \ref{fig:visons-hor}, for comparison to Fig.~\mainfigcorr\ of the Main Text. The dimer-dimer and vison correlators along these other paths show the same type of behavior as for the path shown in the Main Text.

\FloatBarrier
\section{Power-law decay at the critical point}

\begin{figure}[htb]
\includegraphics[width=.33\textwidth]{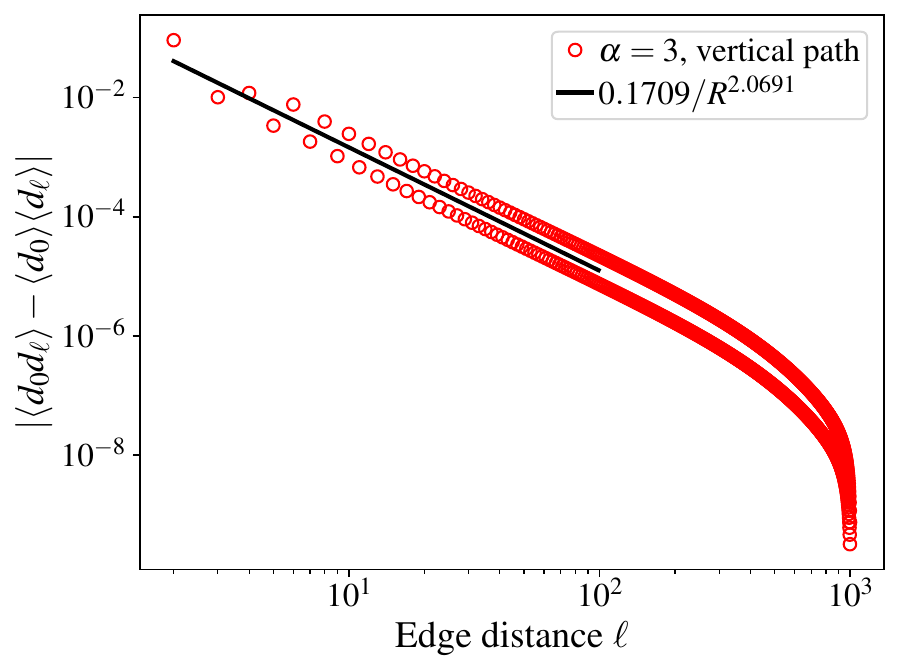}
\includegraphics[width=.33\textwidth]{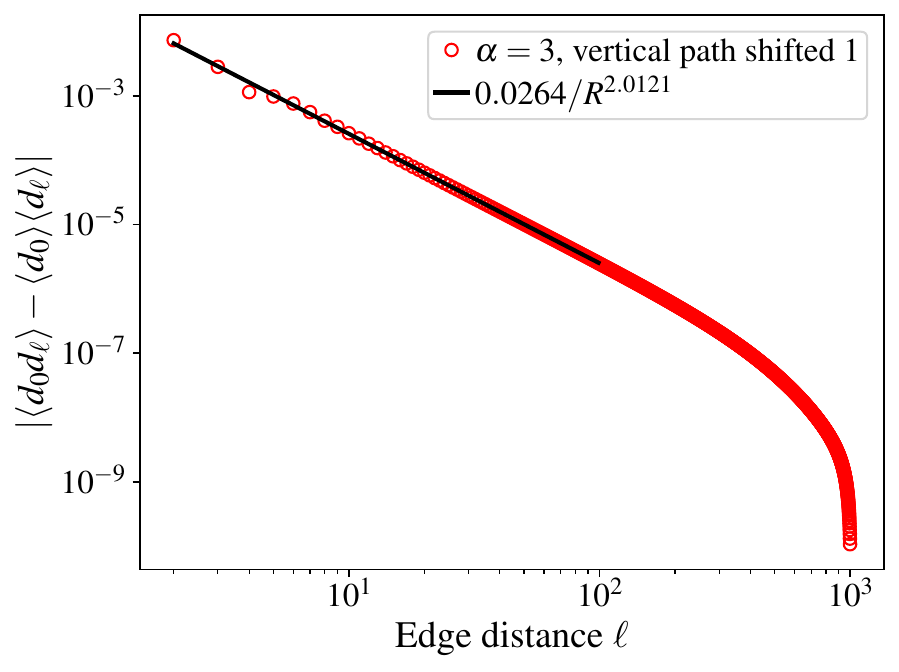}
\includegraphics[width=.33\textwidth]{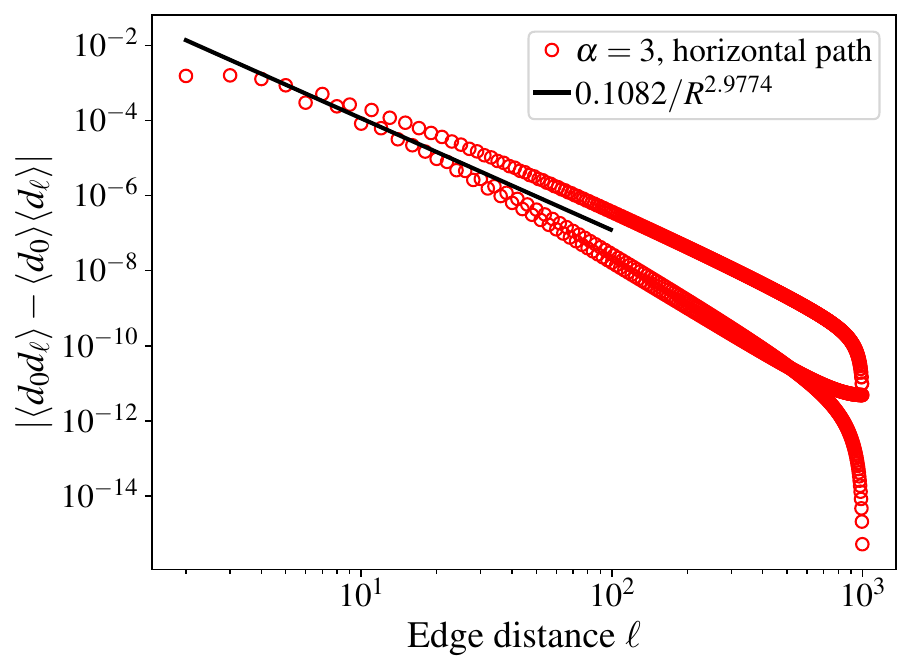}
\caption{Dimer-dimer correlator decay on different paths for the triangular lattice on a size $1001\times1001$ grid with $\wt=3$. All plots use a log-log scale. The best fit curves $a_1/R^{a_2}$ are calculated over distances up to 100. For the vertical paths, the decay is power law $\approx C_2/R^{2}$. For the horizontal path, there appear to be different rates of decay (faster decay than $1/R^2$ but still power-law) depending on the path index.}\label{fig:dimerloglog}
\end{figure}

In this section, we discuss numerics for the dimer-dimer and vison correlators at the critical point $\wt=3$.
We see numerically in \cref{fig:dimerloglog} the dimer-dimer correlator for $\wt=3$ decays as a power-law, although along the horizontal path it appears to decay at a faster power than $1/R^2$. 
Note that for the square lattice at criticality, it is also known that certain pairs have dimer-dimer correlators that can decay as a power-law faster than $1/R^2$ \cite{fisher1963statistical,kenyon2009lectures}.

\begin{figure}[htb]
\includegraphics[width=.33\textwidth]{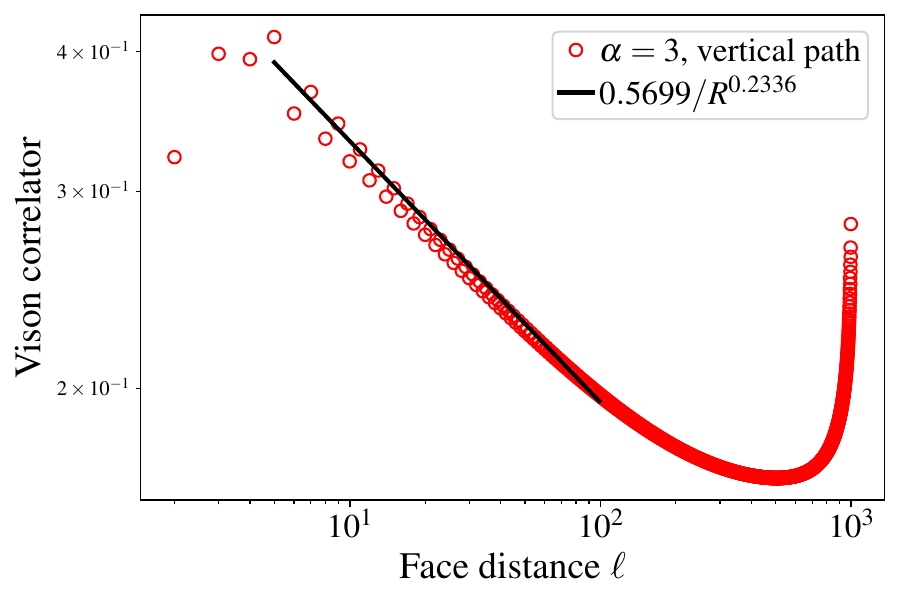}
\includegraphics[width=.33\textwidth]{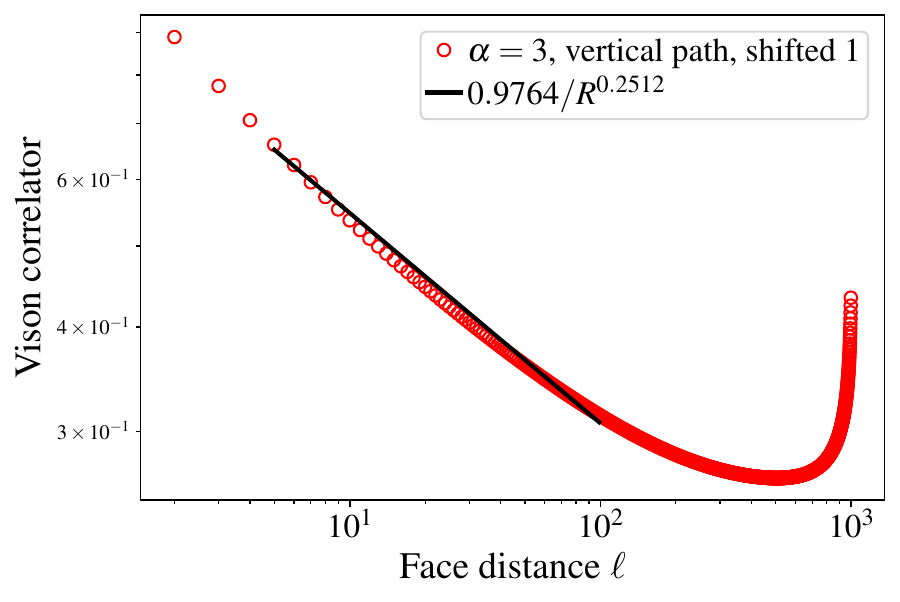}
\includegraphics[width=.33\textwidth]{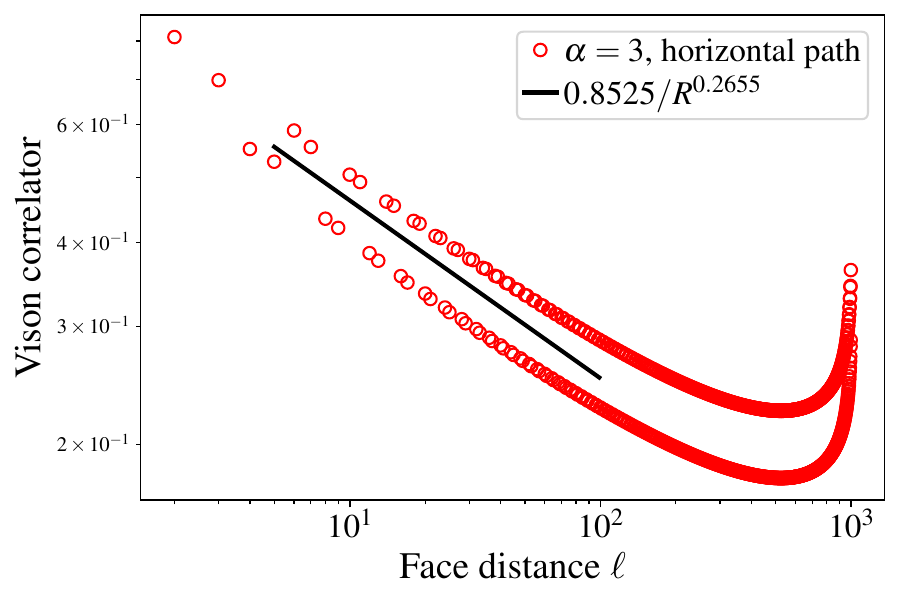}
\caption{Vison correlator decay at the critical value $\wt=3$ for the triangular lattice on a size $1001\times1001$ grid. All plots use a log-log scale. The exponent $\eta$ in the best-fit curve $c/R^\eta$ is somewhat sensitive to the precise region over which the curve fitting is done (here we use 5 to 100), but a value of $\eta=1/4$ is plausible.}\label{fig:visonloglog}
\end{figure}

The vison correlator $|\langle v_0v_\ell\rangle|$ also decays as a power-law at criticality (Fig.~\ref{fig:visonloglog}), though determining the critical exponent $\eta$ in the decay is difficult with only finite-size numerics. It is however plausible that one could have $\eta=1/4$ in agreement with the 2D Ising critical exponent, but we note the numerical values can vary depending on the region used for the linear regression. We anticipate that $\eta$ can be determined more reliably using the infinite-size limit integrals in e.g. \cref{eqn:single-int}.

\FloatBarrier

\section{Critical exponents, finite-size scaling, and curve collapse}

\begin{figure}[htb]
\includegraphics[width=.4\textwidth]{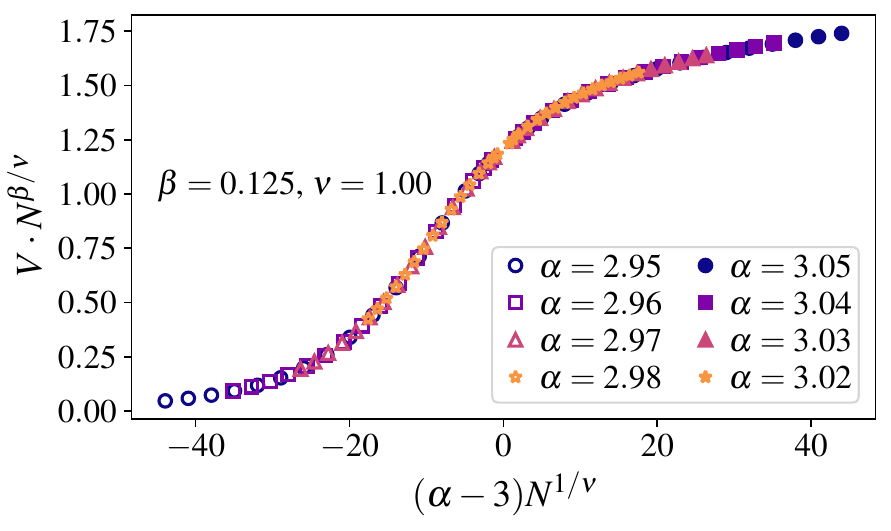}
\includegraphics[width=.4\textwidth]{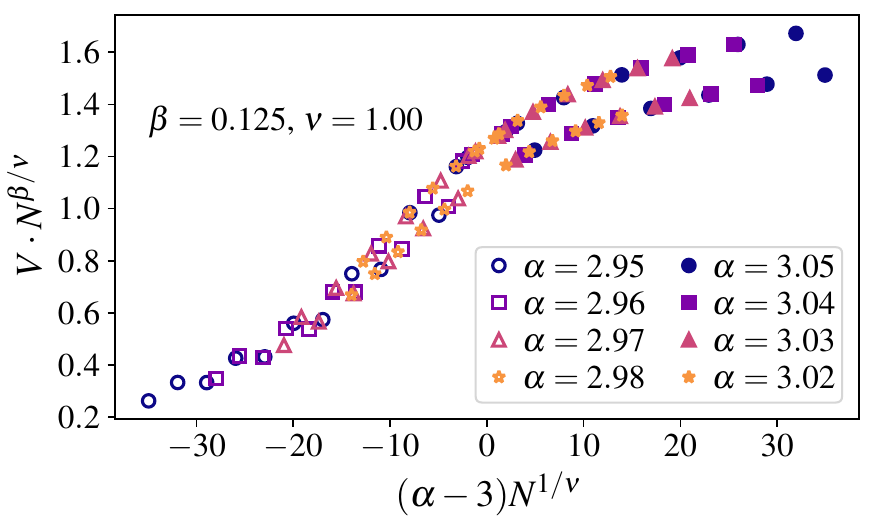}
\caption{(Left) Curve collapse at $\beta=1/8$ and $\nu=1$ for the vertical path through weight $1$ edges, for $N=39,63$ and $99$ through $880$ by $60$s. (Right) Curve collapse at $\beta=1/8$ and $\nu=1$ for horizontal paths starting in a weight 1 face, for $N=39,63$, and $100$ through $700$ by $60$s. The right figure shows two distinct curves since the path endpoint can be in a weight 1 face or weight $\wt$ face, depending on the value of $N$. One could restrict to just a single curve by choosing the values of $N$ carefully or by shifting the path endpoint to always end in the same type of face.}\label{fig:fs_scaling2}
\end{figure}

In this section we consider the finite-size scaling shown in Fig.~\mainfsscaling\ of the Main Text along other lattice paths. There are several different types of paths:
\begin{enumerate}
    \item Vertical path through weight $\wt$ horizontal edges (curve collapse shown in Main Text Fig.~\mainfsscaling)
    \item Vertical path through weight $1$ horizontal edges
    \item Horizontal path starting and ending in weighted faces
    \item Horizontal path starting and ending in unweighted faces
    \item Horizontal path starting and ending in two differently weighted faces
\end{enumerate}
In Fig.~\ref{fig:fs_scaling2}, we consider several of these additional paths and see the curve collapse also occurs for $\beta=1/8$ and $\nu=1$, the same as for the first case shown in the Main Text.

We also comment why we take a square root in the finite-size scaling of $V=\sqrt{|\langle v_0v_\ell\rangle|}$. For magnetization in the Ising model, one has $\lim_{R\to\infty}\langle m(0)m(R)\rangle=\langle m\rangle^2$, where $\langle m\rangle$ is the average magnetization and is the order parameter. Since the critical exponent $\beta$ corresponds to the behavior of the order parameter rather than that of the correlator, we take the square root of the vison correlator $|\langle v_0v_\ell\rangle|$.

\begin{figure}[htb]
\includegraphics[width=.5\textwidth]{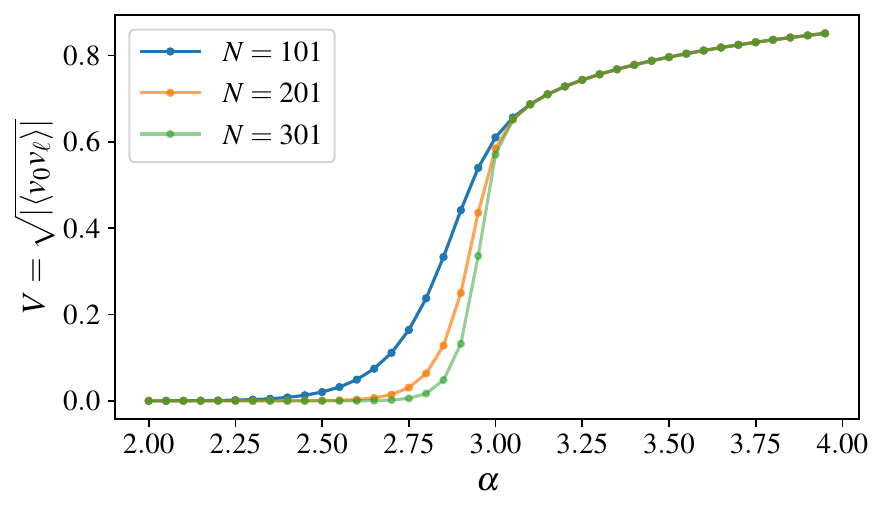}
\caption{Plot of $V=\sqrt{|\langle v_0v_\ell\rangle|}$ calculated from center halfway to the edge, for different system sizes $N=101,201,301$. The curve resembles the magnetization curve for 2D Ising model.}
\label{fig:magnetization}
\end{figure}

For another comparison to the 2D Ising model, in Fig.~\ref{fig:magnetization} we plot $V=\sqrt{|\langle v_0v_\ell\rangle|}$ as a function of the weight $\wt$, and observe the resulting curve closely resembles the magnetization curve for the 2D Ising model.

\FloatBarrier

\section{Double-dimer coverings and constant visons}

In this section we discuss how a phase with only small loops in the double-dimer covering is related to constant vison correlator.

Let $Z=\sum_C \cw(C)$ be the classical partition function, where the sum is over all dimer coverings $C$ and $\cw(C)$ is the classical weight assigned to the covering.
Recall the vison correlator between two faces $f_0,f_1$ is
\begin{align}
\langle f_0f_1\rangle&=\langle(-1)^{\#\text{ dimers along }\gamma}\rangle=\frac{\sum_{C}(-1)^{\#\text{ dimers along }\gamma}\cw(C)}{Z},
\end{align}
where $\gamma$ is a path between $f_0$ and $f_1$. We will consider double-dimer coverings $C\cup C'$, which consist of the edges of two independent random dimer coverings $C$ and $C'$. We can write the vison correlator squared as
\begin{align*}
|\langle f_0f_1\rangle|^2&=\frac{\sum_{C,C'}(-1)^{\#\text{ dimers in $C\cup C'$ along $\gamma$}}\cw(C)\cw(C')}{Z^2}\\
&=\E_{C,C'}[(-1)^{\#\text{ dimers in $C\cup C'$ along $\gamma$}}]\numberthis,
\end{align*}
where we use the notation $\E_{C,C'}$ to represent the average over pairs of independent dimer configurations $C,C'$.
Recalling the edges in $C\cup C'$ form only closed loops and double edges, the parity of $\#\{\text{dimers in $C\cup C'$ along $\gamma$}\}$ is unchanged if the path $\gamma$ enters and exits such a loop or crosses a double edge. So the parity is only affected by loops which contain exactly one of the faces $f_0,f_1$.
For a face $f$, let $L(f)=\#\{\text{loops in $C\cup C'$ that go around $f$}\}$. Then letting $\P_{C,C'}$ denote the probability over pairs of independent dimer configurations $C,C'$,
\begin{align*}
|\langle f_0f_1\rangle|^2&=2\P_{C,C'}[\#\{\text{dimers in $C\cup C'$ along $\gamma$}\}\text{ is even}]-1\\
&=2\P_{C,C'}[L(f_0)\text{ and }L(f_1)\text{ have the same parity}]-1.\numberthis
\end{align*}

If the loops in $C\cup C'$ are all small, such as seen in the ordered phase in the right-mode part of Fig.~\maindouble, then this suggests that $L(f)$ should be a local quantity since it only involves nearby dimers in the small loops.
Then $L(f_0)$ and $L(f_1)$ should be nearly independent for far apart faces due to exponential decay of dimer correlations, giving
\begin{align*}
|\langle f_0f_1\rangle|^2&=2(\P[L(f_0),L(f_1)\in 2\N]+\P[L(f_0),L(f_1)\not\in 2\N])-1\\
&\approx 2\P[L(f_0)\in2\N]\P[L(f_1)\in2\N]+2\P[L(f_0)\not\in2\N]\P[L(f_1)\not\in2\N]-1+O(e^{-cn}).\numberthis
\end{align*}
Letting $p_0=\P[L(f_0)\in2\N]$ and $p_1=\P[L(f_1)\in2\N]$, this becomes
\begin{align*}
|\langle f_0f_1\rangle|^2&\approx 2(p_0p_1+(1-p_0)(1-p_1))-1+O(e^{-cn})\\
&=4p_0p_1-2p_0-2p_1+1+O(e^{-cn}).\numberthis
\end{align*}
To have this tend to $0$ as the distance between the faces goes to $\infty$, we need $p_0=1/2$ or $p_1=1/2$.
If we want the vison correlator to to go to $0$ for all paths, then by using periodicity to consider $f_0$ and $f_1$ the same type of face, we must have $p:=\P[L(f)\in2\N]=1/2$ for all faces $f$ (in the infinite size limit). Due to gauge equivalence of edge weights \cite{kenyon2008height,boutillier2023minimal}, non-uniform edge weights is equivalent to non-uniform face weights. If we have different face weights, or even different face shapes for more complicated lattices, it seems unlikely that we could have all faces have the exact same probability $p_0=1/2$. 
For example, with varying face weights, some faces are likely to have a different average number of edges around them than other faces.
Additionally, the condition $p_0=p_1=1/2$ seems very unstable to perturbations.
Overall, the above suggests that we will not have a topological spin liquid phase or vison decay if the double-dimer coverings have only small loops.

We conjecture, due to the square-octagon example \cite{shah2025breakdown} and related general double-dimer properties of bipartite lattices \cite{kenyon2006dimers}, that it may then be the case that periodic bipartite lattices cannot host a $\Z_2$ quantum spin liquid phase.
We emphasize this is a different suggested reasoning than the conventional argument for lack of spin liquid at the RK point on bipartite lattices \cite{henley1997relaxation,fradkin2004bipartite,moessner2011introduction}, which argues that such (unweighted) models should be critical. However, this cannot hold for bipartite lattices in general, as the unweighted square-octagon lattice is bipartite but exhibits exponential decay of dimer-dimer correlators.

\section{Topological min-entropy transition}

Topological entanglement entropy \cite{kitaev2006topological,levin2006detecting} is used to detect topological order in many-body systems. 
In the setting of quantum dimer models \cite{furukawa2007topological,papanikolaou2007topological,stephan2012renyi,selem2013entanglement}, one considers the set of bonds of the graph $G$, each of which may be occupied by a dimer. The graph $G$ is divided into two subsystems $A$ and $B$, each consisting of a set of bonds. 
Given a state $|\psi\rangle$, such as the ground state of a quantum dimer Hamiltonian, one can consider the reduced density matrix for the subsystem $A$, defined as $\rho_A=\Tr_B|\psi\rangle\langle\psi|$.
The R\'enyi-$p$ entropy of $\rho_A$ is
\begin{align}\label{eqn:renyi}
S_p=\frac{1}{1-p}\log\Tr\rho_A^p,
\end{align}
and the limit $p\to1$ gives the von Neumann entropy
\begin{align}
S_1=-\Tr \rho_A\log\rho_A.
\end{align}
These quantities give a measure of the entanglement between subsystems $A$ and $B$.
For 2D systems with short-range correlations, the entropies $S_p(L)$ generally scale according to the length $L$ of the boundary of $A$.
While the rate of linear scaling is non-universal, for smooth boundaries it was argued in \cite{kitaev2006topological,levin2006detecting} that the next subleading-order constant term in $S_1(L)$ is a \emph{topological} term $\gamma=\log \mathcal D$, where $\mathcal D$ is the total ``quantum dimension'' of the phase, corresponding to $\mathcal D=2$ for a $\Z_2$ spin liquid. There is then the expansion
\begin{align}
S_1(L)=cL-\gamma+\cdots,
\end{align}
where the linear term factor $c$ is non-universal, the terms in $\cdots$ are $o(1)$ as $L\to\infty$, and $\gamma$ is called the \emph{topological entanglement entropy} (TEE).
It was subsequently argued in \cite{flammia2009topological} that the topological R\'enyi entropies $s_p$, defined in an analogous way via
\begin{align}
S_p(L)=c'L-s_p+\cdots,
\end{align}
are also the same, $s_p=\log\mathcal D$, for all non-chiral topological phases.
In practice, it can be difficult to extract the TEE or topological R\'enyi entropies, due to constant-order contributions to $S_p(L)$ from finite-size and boundary effects \cite{kitaev2006topological,levin2006detecting,furukawa2007topological,papanikolaou2007topological,stephan2012renyi}.
However, if one can determine the TEE or a topological R\'enyi entropy, then it can be used to identify topological vs non-topological phases.

TEE and topological R\'enyi entropies have been studied in RK quantum dimer models, where at the RK point the exactly-solvable ground state $|\psi_\mathbf{w}\rangle$ allows one to calculate the entropies $S_p(L)$ in terms of $2^L$ classical dimer probabilities \cite{furukawa2007topological,stephan2012renyi}. 
In particular, the eigenvalues $p_i$ of the reduced density matrix $\rho_A$ are given by classical boundary configuration probabilities.
Additionally, \cite{stephan2012renyi} used an infinite cylinder geometry to extract topological entropies without introducing constant boundary terms, thereby avoiding the need for the boundary-subtraction methods of \cite{kitaev2006topological,levin2006detecting}.

\begin{figure}[htb]
\includegraphics{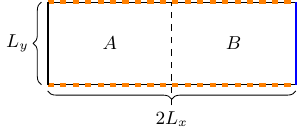}
%\begin{tikzpicture}[scale=.7]
%\def\cwidth{6cm};
%\def\subs{3cm};
%\draw (0,0) rectangle (\cwidth,2);
%\draw[line width=2pt,orange,dashed] (0,2)--++(\cwidth,0);
%\draw[line width=2pt,orange,dashed] (0,0)--++(\cwidth,0);
%\draw[line width=1pt] (0,0)--++(0,2);
%\draw[line width=1pt,blue] (\cwidth,0)--++(0,2);
%\draw[dashed] (\subs,-.5)--++(0,2.5);
%\node at (1.5,1) {$A$};
%\node at (4.5,1) {$B$};
%\draw[decoration={brace,raise=3pt,aspect=.5,amplitude=4pt,mirror=true},decorate] (0,0)--(\cwidth,0);
%\node[below] at (\subs,-.5) {$2L_x$};
%\draw[decoration={brace,raise=3pt,aspect=.5,amplitude=4pt},decorate] (0,0)--(0,2);
%\node[left] at (-.3,1) {$L_y$};
%\end{tikzpicture}
\caption{Cylindrical domain separated into two subsystems $A$ and $B$. The orange dashed lines are identified together.}\label{fig:cyl}
\end{figure}

In this section, we follow the method of \cite{stephan2012renyi} to analytically extract the topological R\'enyi entropy of order $\infty$ (topological min-entropy) for the ground state $|\psi_\mathbf{w}\rangle$ of the $2\times1$-periodic quantum dimer model considered in this paper. The R\'enyi entropy of order $\infty$, or the min-entropy, is the limit of \eqref{eqn:renyi} as $p\to\infty$, 
\begin{align}
S_\infty(L)=-\log p_\mathrm{max}=c'L-s_\infty+\cdots,
\end{align}
where $p_\mathrm{max}$ is the largest of the eigenvalues of $\rho_A$, i.e. in this setting the largest of the classical boundary configuration probabilities. The topological min-entropy (in the absence of constant-order finite-size or boundary terms) is extracted as the subleading term $s_\infty$.
We consider the cylinder geometry shown in Fig.~\ref{fig:cyl}, with radial circumference $L_y$ and with a vertical subsystem cut halfway along the width $2L_x\to\infty$. 
For this cylinder geometry, \cite{stephan2012renyi} analytically showed that the topological min-entropy for the quantum dimer model on the triangular lattice with diagonal edge weight $t>0$, which is in a $\Z_2$ spin liquid phase for all $t>0$, is $s_\infty=\log2$, in agreement with the general theory \cite{kitaev2006topological,levin2006detecting}.

Here we analytically show that for the $2\times1$ periodic triangular lattice with one horizontal edge weight $\wt$ and other edge weights one,
\begin{align}
s_\infty=\begin{cases}\log2,&\wt<3\\
0,&\wt>3
\end{cases}.
\end{align} 
This analytically demonstrates the topological nature of the phase transition at $\wt=3$, from the topological $\Z_2$ spin liquid phase to a topologically trivial phase.

\begin{center}
\textbf{A. Calculation on a cylinder}
\end{center}

\begin{figure}[htb]
\includegraphics{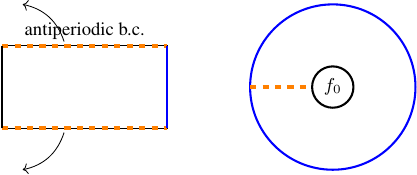}
%\begin{tikzpicture}[scale=.7]
%\draw (0,0) rectangle (4,2);
%\draw[line width=2pt,orange,dashed] (0,2)--++(4,0);
%\draw[line width=2pt,orange,dashed] (0,0)--++(4,0);
%\draw[line width=1pt] (0,0)--++(0,2);
%\draw[line width=1pt,blue] (4,0)--++(0,2);
%\node[above] at (2,2) {antiperiodic b.c.};
%\draw[->] (1.5,-.1) to [bend left] (.5,-1);
%\draw[->] (1.5,2.1) to [bend right] (.5,3);
%\begin{scope}[xshift=8cm]
%\draw[line width=1pt,blue] (0,1) circle (2cm);
%\draw[line width=1pt] (0,1) circle (.5cm);
%\draw[line width=2pt,orange,dashed](-2,1)--++(1.5,0);
%\node at (0,1) {$f_0$};
%\end{scope}
%\end{tikzpicture}
\caption{Embedding a cylinder in the plane. Due to the parity of our choice of height $L_y\in2\N$ and the clockwise-odd rule for the new face $f_0$, we need to assign antiperiodic boundary conditions on cylinder since there are an even number of edges in the boundary of $f_0$.}\label{fig:embed}
\end{figure}

As in \cite{stephan2012renyi}, on the cylinder we generally have~\footnote{Formally, we do not prove the second equality, though it holds intuitively and numerically as in \cite{stephan2012renyi}.}
\begin{align*}
S_\infty(L)&=-\log p_\mathrm{max}
=-\log\left[\frac{Z_{\mathrm{cyl}}(L_x/2,L_y)^2}{Z_{\mathrm{cyl}}(L_x,L_y)}\right],\numberthis\label{eqn:S-cylinder}
\end{align*}
and will extract the subleading constant $s_\infty$ in $S_\infty(L)=cL-s_\infty+\cdots$ by evaluating \eqref{eqn:S-cylinder} using Kasteleyn/Pfaffian methods.
We take the horizontal cylinder in Fig.~\ref{fig:cyl} rather than the vertical one in \cite{stephan2012renyi}, which will simplify some calculations.
Kasteleyn's method for computing the classical partition function can be used on the cylinder by embedding the cylinder in the plane as in Fig.~\ref{fig:embed}.
Due to the choice of height $L_y\in2\N$ (here meaning there are $L_y$ edges around the shortest loop around the cylinder), we will have to take antiperiodic boundary conditions in the Kasteleyn matrix for the edges which wrap around the dotted orange identification.
Then counting dimer configurations on the cylinder only requires a single Kasteleyn matrix.

\begin{figure}[htb]
\includegraphics{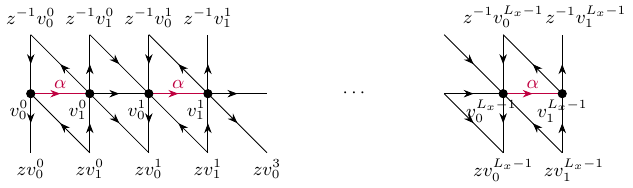}
\caption{Fundamental domain for the cylinder of fixed width $2L_x\in2\N$ and height $L_y$, with the top and bottom edges identified. The vertex $v^i$ represents the translate of vertex $v=v^0$ by translation $T_{(2,0)}$ in the $x$-direction.}\label{fig:fdcyl}
\end{figure}

Evaluation of \eqref{eqn:S-cylinder} starts similarly to evaluation on the torus, but with only a block Fourier transform in the periodic $y$-direction.
Since we consider antiperiodic boundary conditions, let $k_x=(2x+1)\pi/L_x$ and $k_y=(2y+1)\pi/L_y$, and also define
$w_x=e^{i\pi(2x+1)/L_x}=e^{ik_x}$ and $z_y=e^{i\pi(2y+1)/L_y}=e^{ik_y}$.
Let $K$ be the Kasteleyn matrix for the cylinder domain with antiperiodic boundary conditions in the $y$-direction.
We consider the fundamental domain shown in Fig.~\ref{fig:fdcyl}. Note that the width is $2L_x$ since we allow $L_x$ horizontal translations of a $2\times1$ domain.
Our goal is to calculate $\det K$ to obtain the partition function $Z_{\mathrm{cyl}}(L_x,L_y)=\sqrt{\det K}$.
Using translation invariance under the operator $R_{(0,1)}$, which represents translation by $(0,1)$ followed by negation of coordinates in the first row,
we can block diagonalize the Kasteleyn matrix in this direction to get $L_y$ blocks $K^{(y)}$ each of size $2L_x\times 2L_x$. Each $K^{(y)}$ is given the Kasteleyn matrix for the fundamental domain in Fig.~\ref{fig:fdcyl} with $z=z_y$, and we then have
\begin{align}\label{eqn:detK1}
\det K&=\prod_{y=0}^{L_y-1}\det K^{(y)}.
\end{align}

The matrix $K^{(y)}$ is not easy to further block-diagonalize, due to the non-periodic boundary conditions in the $x$-direction.
However, one can add horizontal wrap-around edges to vertices $v_0$ (left-most) and $v_{1}^{L_x-1}$ (right-most) in Fig.~\ref{fig:fdcyl} to form a new matrix $K_0^{(y)}$ which is the same as $K^{(y)}$ except it adds two entries for the connections between $v_0$ and $v_1^{L_x-1}$ with aperiodic boundary conditions,
\begin{align}\label{eqn:k0y}
K_0^{(y)}|_{\{v_0,v_1^{L_x-1}\}}&=\begin{bmatrix}\langle v_0|K^{(y)}|v_0\rangle&1+z_y\\
-1-z_y^{-1}&\langle  v_1^{L_x-1}|K^{(y)}|v_1^{L_x-1}\rangle
\end{bmatrix}.
\end{align}
(Note there are no variables $w$ since we don't periodicize in that direction; we are simply connecting $v_0$ and $v_1^{L_x-1}$.)

Since $K_0^{(y)}$ is a small perturbation of $K^{(y)}$, we will be able to relate their determinants later.
We can block-diagonalize $K_0^{(y)}$ using that it commutes with the operator $R_{(2,0)}$, which represents translation by $(2,0)$ followed by negation of coordinates in the first two columns. We think of the domain in Fig.~\ref{fig:fdcyl} as consisting of $L_x$ horizontal translates of the torus fundamental domain from Fig.~\ref{fig:fd6}.
Define the block 1D Fourier transform matrix
\begin{align}
\langle u^\ell|W|v^k\rangle&=\begin{cases}\frac{1}{\sqrt{L_x}}e^{-i\pi(2\ell+1)k/L_x},&u=v\\
0,&\text{otherwise}
\end{cases},
\end{align}
and let $L_0^{(y)}=WK_0^{(y)}W^{-1}$. 
Note that $W^{-1}$ is given as $\langle u^\ell|W^{-1}|u^k\rangle=\frac{1}{\sqrt{L_x}}e^{i\pi(2k+1)\ell/L_x}$.
Then
\begin{align*}
\langle u^\ell|L_0^{(y)}|v^k\rangle &= \frac{1}{L_x}\sum_{\ell',k'=0}^{L_x-1}e^{-i\pi(2\ell+1)\ell'/L_x}e^{i\pi(2k+1)k'/L_x}\langle u^{\ell'}|K_0^{(y)}|v^{k'}\rangle\\
&=\sum_{j'=-1,0,1}e^{i\pi(2\ell+1)j'/L_x}\langle u|K_0^{(y)}|v^{j'}\rangle
=\sum_{j'=-1,0,1}w_\ell^{j'}\langle u|K_0^{(y)}|v^{j'}\rangle,\numberthis
\end{align*}
using that $\ell=k$ is needed for the above to be nonzero (since $L_0^{(y)}$ is a block matrix).
The matrix $L_0^{(y)}$ is made up of $2\times2$ blocks, which we can index using coordinates $(v_0^{\ell'},v_1^{\ell'})$ to write each block as
\begin{align}
(L_0^{(y)})_{w_{\ell'},z}&=\begin{bmatrix}
z-z^{-1}&\wt-z-w_{\ell'}^{-1}-z^{-1}w_{\ell'}^{-1}\\
-\wt+z^{-1}+w_{\ell'}+zw_{\ell'}&-z+z^{-1}
\end{bmatrix},
\end{align}
which is the same as \eqref{eqn:L}. The inverse $(L_0^{(y)})_{w,z}^{-1}$ and characteristic polynomial $P(w,z)=\det L_{w,z}$ are then given by 
\begin{align}\label{eqn:L0inv}
(L_0^{(y)})_{w,z}^{-1}=\frac{1}{P(w,z)}\begin{bmatrix}-z+z^{-1}&-\wt+z+w^{-1}+z^{-1}w^{-1}\\
\wt-z^{-1}-w-zw&z-z^{-1}
\end{bmatrix},
\end{align}
and
\begin{align}
P(w,z)=\det L_{w,z}=-(z-z^{-1})^2+(\wt-z-w^{-1}-z^{-1}w^{-1})(\wt-w-z^{-1}-zw),
\end{align}
as before.
Since $(K_0^{(y)})^{-1}=W^{-1}(L_0^{(y)})^{-1}W$, we have
\begin{align*}
\langle u^\ell|(K_0^{(y)})^{-1}|v^k\rangle&=\frac{1}{L_x}\sum_{\ell',k'=0}^{L_x-1}e^{i\pi(2\ell'+1)\ell/L_x}e^{-i\pi(2k'+1)k/L_x}\langle u^{\ell'}|(L_0^{(\ell)})^{-1}|v^{k'}\rangle\\
&=\frac{1}{L_x}\sum_{\ell'=0}^{L_x-1}w_{\ell'}^{\ell-k}\langle u|(L_0^{(y)})_{w_{\ell'},z_y}^{-1}|v\rangle,\numberthis
\end{align*}
using that $L_0^{(\ell)}$ is a block matrix so $u^{\ell'}$ and $v^{k'}$ must belong to the same block, $\ell'=k'$. 
We then have for example
\begin{align*}
\langle v_0|(K_0^{(y)})^{-1}|v_1^{L_x-1}\rangle&=\frac{1}{L_x}\sum_{\ell'=0}^{L_x-1}e^{-i\pi(2\ell'+1)(L_x-1)/L_x}\langle v_0|(L_0^{(y)})^{-1}_{w_{\ell'},z_y}|v_1\rangle\\
&=\frac{1}{L_x}\sum_{\ell'=0}^{L_x-1}-w_{\ell'}\langle v_0|(L_0^{(y)})^{-1}_{w_{\ell'},z_y}|v_1\rangle.
\end{align*}
Considering the restriction to the end vertices $v_0$ and $v_1^{L_x-1}$, using \eqref{eqn:L0inv} shows
\begin{align}
(K_0^{(y)})^{-1}_{\{v_0,v_1^{L_x-1}\}}&=\frac{1}{L_x}\sum_{\ell'=0}^{L_x-1}
\frac{1}{P(w_{\ell'},z_y)}
\begin{bmatrix}
-z_{y}+z_{y}^{-1}&\wt w_{\ell'}-z_yw_{\ell'}-1-z_y^{-1} \\
-\wt w_{\ell'}^{-1}+z_y^{-1}w_{\ell'}^{-1}+1+z_y&z_{y}-z_{y}^{-1}
\end{bmatrix}\\
&=:\begin{bmatrix}
A_{11}^{(y)}&A_{12}^{(y)}\\
-\bar A_{12}^{(y)}&-A_{11}^{(y)}
\end{bmatrix}.
\end{align}
Since $K^{(y)}=K_0^{(y)}+(K^{(y)}-K_0^{(y)})$, then
\begin{align}
\frac{\det K^{(y)}}{\det K_0^{(y)}}&=\det(I+(K_0^{(y)})^{-1}(K^{(y)}-K_0^{(y)}))=:\det M^{(y)}.
\end{align}
As $K^{(y)}-K_0^{(y)}$ has only two nonzero entries [see \eqref{eqn:k0y}],
we can consider the nonzero $2\times2$ part of $M^{(y)}$, which we will call $M_2^{(y)}$:
\begin{align*}
M_2^{(y)}&=I_2+(K_0^{(y)})^{-1}_{\{0,L_x-1\}}\begin{bmatrix}0&-1-z_y\\1+z_y^{-1}&0\end{bmatrix}\\
&=\begin{bmatrix}
1+A_{12}^{(y)}(1+z_y^{-1}) & A_{11}^{(y)}(1+z_y) \\
A_{11}^{(y)}(1+z_y^{-1})&1+\bar A_{12}^{(y)}(1+z_y)
\end{bmatrix}.\numberthis
\end{align*}
Using that $A_{11}^{(y)}$ is purely imaginary, we thus obtain
\begin{align}\label{eqn:detM2}
\det M_2^{(y)}
&=|1+A_{12}^{(y)}(1+e^{-ik_y})|^2+|A_{11}^{(y)}|^2|1+e^{ik_y}|^2.
\end{align}
Since $\det K^{(y)}=\det M_2^{(y)}\det K_0^{(y)}$, using \eqref{eqn:detK1} we see
\begin{align}
\frac{Z_{\mathrm{cyl}}(L_x/2,L_y)^2}{Z_{\mathrm{cyl}}(L_x,L_y)}&=\frac{\prod_{y=0}^{L_y-1}\det M_2^{(y)}\prod_{x=0}^{L_x/2-1}P(e^{i\pi(2x+1)2/L_x},e^{i\pi(2y+1)/L_y})}{\prod_{y=0}^{L_y-1}(\det M_2^{(y)})^{1/2}\prod_{x=0}^{L_x-1}P(e^{i\pi(2x+1)/L_x},e^{i\pi(2y+1)/L_y})^{1/2}}.
\end{align}
We consider the non-critical case only, i.e. $\wt\ne3$, so that $P$ has no zeros on $S^1\times S^1$.
Then taking the limit $L_x\to\infty$, the ratios involving the characteristic polynomial $P$ tend to $1$ using the Euler--Maclaurin formula, which then gives for $L=L_y$
\begin{align}
S_\infty(L)&=-\frac{1}{2}\sum_{y=0}^{L-1}\log\left[\lim_{L_x\to\infty}\det M_2^{(y)}\right]\nonumber\\
&=-\frac{1}{2}\sum_{k_y=(2y+1)\pi/L}^{0\le y\le L-1}\log\left[\lim_{L_x\to\infty}|1+A_{12}^{(y)}(1+e^{-ik_y})|^2+|A_{11}^{(y)}|^2|1+e^{ik_y}|^2\right].\label{eqn:Sinf-sum}
\end{align}
Taking the limit $L_x\to\infty$ using \eqref{eqn:single-int} gives
\begin{align}
A_{11}^{(y)}&=\frac{1}{2\pi i}\oint_{S^1}\frac{-z+z^{-1}}{P(w,z)}\,\frac{dw}{w}
=\frac{-2i\sin(k_y)}{\sqrt{c_2(e^{ik_y},\wt)^2-4c_1(e^{ik_y},\wt)c_3(e^{ik_y},\wt)}},\label{eqn:A11}\\
A_{12}^{(y)}&=\frac{1}{2\pi i}\oint_{S^1}\frac{(\wt-z)w-(1+z^{-1})}{P(w,z)}\,\frac{dw}{w}
=\frac{(\wt-e^{ik_y}) r_+(e^{ik_y},\wt)-(1+e^{-ik_y})}{\sqrt{c_2(e^{ik_y},\wt)^2-4c_1(e^{ik_y},\wt)c_3(e^{ik_y},\wt)}},
\end{align}
where $c_1,c_2,c_3,r_+$ are as in Section~\ref{sec:expdecay}.
To evaluate \eqref{eqn:Sinf-sum} in the limit $L=L_y\to\infty$, we want to use the Euler--Maclaurin formula [\eqref{eqn:em2} below] to replace the sum with an integral. This works as long as the integrand has no singularities, i.e. as long as $\det M_2^{(y)}>0$ in the $L_x\to\infty$ limit. From the second term in \eqref{eqn:detM2} and from \eqref{eqn:A11}, we can only possibly have $\det M_2^{(y)}=0$ if $k_y=0$ or $k_y=\pi$. If $k_y=\pi$ then $\det M_2^{(y)}=1$, so we only need to consider if there is a zero at $k_y=0$.

As $k_y\to0$,
\begin{align}
A_{11}^{(y)}&\xrightarrow{k_y\to0}\begin{cases}
0,&\wt\ne3\\
\pm\frac{i}{4},&\wt=3
\end{cases},\qquad
A_{12}^{(y)}\xrightarrow{k_y\to0}\begin{cases}-\frac{1}{2},&a<3\\
-\frac{1}{4},&a=3\\
0,&a>3
\end{cases},
\end{align}
using
\begin{align*}
c_1(1,\wt)&=c_3(1,\wt)=2-2\wt,\quad c_2(1,\wt)=5+\wt^2-2\wt,
\quad\sqrt{c_2^2-4c_1c_3}=|(\wt+1)(\wt-3)|,\\
r_+(1,\wt)
&=\frac{-5-\wt^2+2\wt+|(\wt+1)(\wt-3)|}{4-4\wt}
=\begin{cases}
\frac{2}{\wt-1},&\wt\ge3\\
\frac{\wt-1}{2},&\wt<3
\end{cases}.
\end{align*}
Therefore,
\begin{align}
\lim_{k_y\to0}\det M_2^{(y)}&=|1+2A_{12}|^2+4|A_{11}|^2
=\begin{cases}0,&\wt<3\\
1/2,&\wt=3\\
1,&\wt>3
\end{cases}.
\end{align}
For $\wt>3$, there are then no singularities in \eqref{eqn:Sinf-sum}, and we immediately get $s_\infty=0$ using the Euler--Maclaurin formula calculation in \eqref{eqn:em2}.
For $\wt<3$, we have a singularity as $k_y\to0$, so we need to subtract out the singularity to apply the Euler--Maclaurin formula.
Let
\begin{align*}
f(y)&=-\frac{1}{2}\log\left[|1+A_{12}^{(y)}(1+e^{-ik_y})|^2+|A_{11}^{(y)}|^2|1+e^{ik_y}|^2\right].
\end{align*}
We can Taylor expand around $k_y=0$ to find for $\wt<3$,
\begin{align}\label{eqn:taylor0}
|1+A_{12}^{(y)}(1+e^{-ik_y})|^2+|A_{11}^{(y)}|^2|1+e^{ik_y}|^2&=\frac{(3-\wt)^2(1+\wt)^2k_y^2+16k_y^2+O(k_y^3)}{d(e^{ik_y},\wt)^2},
\end{align}
where $d(e^{ik_y},\wt):=\sqrt{c_2(e^{ik_y},\wt)^2-4c_1(e^{ik_y},\wt)c_3(e^{ik_y},\wt)}=(3-\wt)(1+\wt)+O(k_y^2)$.
Then \eqref{eqn:taylor0} is $c_\wt k^2+O(k^3)$, so the divergence of $f$ as $k_y\to0$ is $-\log k$.

Let $g(k):=f(k)+\log k+\log(2\pi-k)$, with the second term added due to periodicity. 
For a smooth $C^\infty$ function $g$ which is periodic on $[0,2\pi]$, the Euler--Maclaurin formula gives, as $L\to\infty$,
\begin{align*}
\sum_{y=0}^{L-1}g\bigg(\frac{(2y+1)\pi}{L}\bigg)&=\int_0^{L-1}g\bigg(\frac{(2y+1)\pi}{L}\bigg)\,dy +\frac{1}{2}\left[g(\pi/L)+g\Big(2\pi-\frac{\pi}{L}\Big)\right]+o(1)\\
&=\frac{L}{2\pi}\int_0^{2\pi}g(k)\,dk-\frac{L}{2\pi}\int_{0}^{\pi/L}g(k)\,dk-\frac{L}{2\pi}\int_{2\pi-\pi/L}^{2\pi}g(k)\,dk+\frac{1}{2}\left[g(\pi/L)+g\Big(2\pi-\frac{\pi}{L}\Big)\right]+o(1)\\
&=\frac{L}{2\pi}\int_0^{2\pi}g(k)\,dk+o(1).\numberthis\label{eqn:em2}
\end{align*}
Note there is no subleading constant-order term.

It remains to consider the sum over $-\log k_y-\log(2\pi-k_y)$.
Using Stirling's formula, the contribution from the $-\log k_y$ is
\begin{align*}
-\sum_{y=0}^{L-1}\log\left(\frac{(2y+1)\pi}{L}\right)&=-L(\log(2\pi) -1)-\frac{1}{2}\log 2+O(1/L),
\end{align*}
which gives a subleading constant term $-\frac{1}{2}\log2$.
We obtain another such factor from $\log(2\pi-k)$, which gives total subleading constant term $-\log 2$ in \eqref{eqn:Sinf-sum} as $L_y\to\infty$.
We thus obtain
\begin{align}
s_\infty&=\begin{cases}\log2,&\wt<3\\
0,&\wt>3
\end{cases},
\end{align}
as claimed.

\begin{rmk}
Finite-size numerical calculation of classical boundary configuration probabilities $p_i$ confirms the above analytical results for the topological min-entropy $s_\infty$, and also gives the TEE behavior $\gamma=\log2$ for $\wt<3$, and $\gamma=0$ for $\wt>3$.
\end{rmk}

\bibliography{dimers}